\documentclass[journal]{IEEEtran}
\usepackage{amsmath,amssymb}
\usepackage{graphicx}
\usepackage{booktabs}
\usepackage{enumitem}
\usepackage[hidelinks]{hyperref}
\graphicspath{{figures/}}
\begin{document}
\title{Passivity Correction of the Thermistor Model in Bolometric Power Sensors and Its Dependence on Bead Temperature}
\author{Patricio~Greco,~\IEEEmembership{Member,~IEEE}%
\thanks{P. Greco is an independent researcher, Buenos Aires, Argentina. ORCID 0000-0002-4255-7558.}}
\markboth{}{Greco: Passivity Correction of the Thermistor Model}
\maketitle

\begin{abstract}
This paper closes a series of works on thermistor bolometric power sensors, coaxial and waveguide, by treating what the two have in common. The per-state transformation $m$ introduced in the waveguide work, which corrects the thermistor impedances of the model in each bridge state so that the two-port of the mount is passive, is analyzed on the eleven sensors of the preceding papers: the corrections are a few tenths of a percent in the coaxial mounts and one to two percent in the waveguide mounts, follow one pattern within each family, and the hottest state is the one whose efficiency departs most from the others. The transformation is tested by an independent measurement: the substituted powers of the three states, read with the manufacturer's procedure against a leveled source whose reflection is obtained from the closure of three Apollonius circles, give efficiency ratios of $0.999\pm0.016$ and $0.997\pm0.008$ ($k=2$), which contain the values at the passivity floor (0.996, 0.998) and exclude those of the first pass (0.941, 1.023). Three sensors were swept in the bridge resistance, 23 and 31 states, and a fourth is treated apart: the correction reduces to one constant per state, the parallel resistance of the bead, represented by a law with one term scaling with the bridge resistance and one exponential in the bead temperature, with one thermal coefficient per family. The same property explains why the cross-ratio of four measured reflection coefficients, invariant under any state-independent two-port, departs from that of the four bridge resistances.
\end{abstract}

\section{Introduction}
\label{sec:intro}
\begin{figure}[!b]
\centering
\includegraphics[width=0.9\columnwidth]{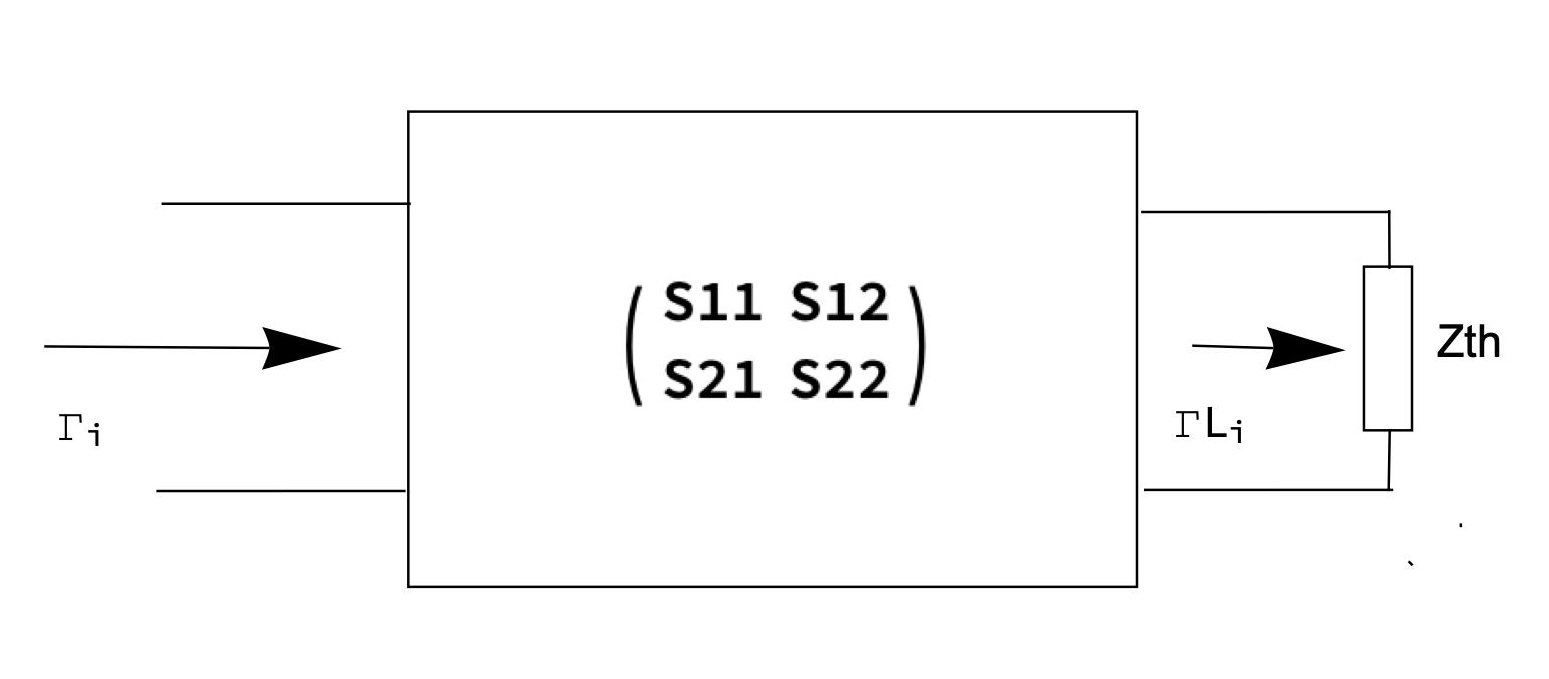}
\caption{Two-port representation of the sensor, as in~\cite{greco2026coax}: input reflection coefficient $\Gamma_i$, body scattering matrix $S$, and thermistor load $Z_{\mathrm{th},i}$ with reflection coefficient $\Gamma_{L,i}$; $i$ denotes the state.}
\label{fig:scheme}
\end{figure}

\IEEEPARstart{I}{n} the course of the investigation of thermistor bolometric power sensors, once an equivalent model and the physical conditions that determine its parameters had been established, with results consistent with the available calorimetric certificates both for the coaxial mounts (478A, 8478B) \cite{greco2026coax} and for the K486 waveguide mounts and one specimen of the R486 \cite{greco2026wg}, all the sensors examined---regardless of their characteristics and of the differences in their physical construction---were found to share one problem: the passivity of the scattering matrix implied by the model. We recall that the basic model, which can be found in either of the reference documents, consists of a thermistor represented by its impedance $Z_{\mathrm{th},i}$, where $i$ denotes the state in which the reflection coefficient of the sensor $\Gamma_i$ is measured, $i=2$ being the central state recommended by the manufacturer for the operation of the sensor, and of a scattering matrix $S$ common to all the states, which represents the physical connection between the thermistor and the sensor port (Fig.~\ref{fig:scheme}).

The efficiency of a bolometer mount has been obtained from impedance data since the work of Kerns~\cite{kerns1949}, Beatty and Reggia~\cite{beatty1955} and Engen~\cite{engen1961,engen1992}, and the microcalorimeter~\cite{macpherson1955} remains the primary standard against which such determinations are judged, in coaxial line~\cite{nist1357} and in waveguide~\cite{nist1511,judaschke2009,cui2016,perangin2023}, with the direct-comparison transfer~\cite{weidman1996,juroshek2000} and the key comparisons of calibration factor~\cite{devreede2005} as the link to the working standards; thermoelectric and bolometric standards have been compared in~\cite{brunetti2013}. The thermistor itself has been modeled as a nonlinear element by Kazemipour and Ziad\'e~\cite{kazemipour2010,ziade2011}, and the two preceding works of this series~\cite{greco2026coax,greco2026wg} determine the calibration factor from the reflection coefficients of three bridge states with a physically constrained model of the bead. Kazemipour and Ziad\'e report a capacitance of the bare thermistor, measured in a test fixture, that depends on the bias resistance; what none of these works describes is how the bead changes within its mount from one bridge state to the next, as a function of its temperature and under the passivity of the mount, nor tests that change by substituted power; the present paper is about that change, and it uses the microcalorimeter literature only as the reference the final verification is compared with.

This passivity problem has been analyzed at length. What could be established is that no fixed set of model parameters common to the three states yields a passive scattering matrix: the parameters that reproduce any one state, when held in the other two, leave the non-passivity at the level of the first pass~\cite{greco2026wg}---the fit of the model and of the two-port to the three reflection coefficients in~\cite{greco2026coax,greco2026wg}, that is, the determination of the sensor from its reflection coefficients alone, before any condition of passivity is imposed, is called here the first pass. The adjustments required to bring the model to passivity must therefore be produced within the set of impedance triples that leave the result of the first pass invariant, which we denote $\mathcal{M}_F$. In this sense the model obtained in the first pass is the representative one: it already contains all the results ordinarily used of a sensor---those expected from a calorimetric calibration certificate---and requires only a correction within $\mathcal{M}_F$ to complete the description of the sensor and of its thermistor.

To restore the passivity of the scattering matrix, per-state corrections are introduced on the impedances $Z_{\mathrm{th},i}$,
\begin{equation}
Z'_i = m_i\,Z_{\mathrm{th},i},\qquad m_i = e^{\,s_i + j\theta_i},\qquad i=1,2,3,
\label{eq:m}
\end{equation}
in such a way that the results obtained in the first pass are preserved. The scattering matrix $S(Z',\Gamma)$ is the two-port determined by the three pairs $(Z'_i,\Gamma_i)$ through the bilinear relation between load and input reflection coefficients; it is passive if and only if its spectral norm satisfies $\sigma_{\max}(S)\le 1$, i.e.\ no excitation extracts more power than it delivers. The paper is written to be read after~\cite{greco2026coax}: the model, its parameters and the determination of the two-port are used here as established there. Each state has a calibration factor, $K_{f,i}=\eta_i(1-|\Gamma_i|^2)$~\cite{greco2026coax}; the calibration factor of the sensor, $K_f=K_{f,2}$, is that of the operating state, and it factorizes as $K_f=F\cdot G$, with $G$ a function of the measured $\Gamma_i$ alone \cite{greco2026coax}, so that the first pass fixes
\begin{equation}
F(z_1,z_2,z_3)=\frac{2\,\mathrm{Re}[z_2]\,|z_3-z_1|}{|z_2-z_1|\,|z_3-z_2|},
\label{eq:F}
\end{equation}
where $z_i=Z_{\mathrm{th},i}$ are the thermistor impedances of the first-pass model~\cite{greco2026coax} ($F$ is homogeneous of degree zero, so the impedances may be taken in ohms or normalized), and the set of impedance triples that share its value,
\begin{equation}
\mathcal{M}_F=\{(z_1,z_2,z_3):\ F(z_1,z_2,z_3)=F_0\},
\label{eq:MF}
\end{equation}
is such that
the calibration factor $K_f$ and the efficiency $\eta_2$ of the operating state are unchanged by any displacement within $\mathcal{M}_F$. The spectral norm, in turn, is bounded from below by the measured efficiency of the operating state, $\sigma_{\max}^2\ge\eta_2$~\cite{greco2026wg}, since the efficiency is a power gain of the two-port and no gain exceeds $\sigma_{\max}^2$ (Appendix~\ref{app:floor}); the passivity floor $\sigma_{\max}=\sqrt{\eta_2}$ is therefore the least active two-port compatible with the data. The correction is selected by bringing $S$ to that floor with the smallest departure of the impedances from the model:
\begin{equation}
\begin{split}
\min_{s,\theta}\ &\sum_{i=1}^{3}\left(s_i^2+\theta_i^2\right)\qquad(\theta_i\ \text{in radians})\\
\text{subject to}\quad &F(Z')=F_0,\qquad \sigma_{\max}\!\left(S(Z',\Gamma)\right)\le\sqrt{\eta_2}+\varepsilon\,.
\end{split}
\label{eq:minnorm}
\end{equation}
The bound $\sqrt{\eta_2}$ is not attained exactly: the least $\sigma_{\max}$ compatible with $F=F_0$ lies above it by at least a term of second order in the reflection coefficients of the operating state, $(1-\eta_2)(|\Gamma_2|^2+\eta_2|\Gamma_{L,2}|^2)/2\sqrt{\eta_2}$, where $\Gamma_{L,2}$ is the reflection coefficient of the thermistor impedance $Z_{\mathrm{th},2}$ at the bead port (Fig.~\ref{fig:scheme}; $Z_{\mathrm{th}}$ is the impedance of the thermistor model, which differs somewhat from one mount model to another~\cite{greco2026coax,greco2026wg}; Appendix~\ref{app:floor}), below $10^{-4}$ in the coaxial mounts and $3\times10^{-4}$ in band median in the waveguide mounts, where it exceeds $10^{-3}$ only at the lower edge of the band (18.0--18.3~GHz of specimen~4, up to $1.4\times10^{-3}$). The margin $\varepsilon$ makes the constraint attainable; it is set at $10^{-3}$ in every sensor, as in~\cite{greco2026wg}, which covers that term except at those edge points. Since the first pass violates the constraint in every sensor ($\sigma_{\max}>1$, Section~\ref{sec:results}), the constraint is active and the selection holds it with equality, $\sigma_{\max}=\sqrt{\eta_2}+\varepsilon$, at every frequency where that value is attainable; at the edge points it settles at the least $\sigma_{\max}$ compatible with $F=F_0$. The least attainable $\sigma_{\max}$ lies between the bound and the margin, and the result does not depend on where the margin is placed (Section~\ref{sec:results}). The same bound holds for every state, $\sigma_{\max}^2\ge\eta_i$ (Appendix~\ref{app:floor}), so that at the floor no state can be more efficient than the operating one by more than $2\times10^{-3}$: the operating state is, by construction, the most efficient state of the sensor within that margin, and what the selection determines is how far below $\eta_2$ the other states lie.

The purpose of this paper is to approach a physical explanation of the meaning of the transformation $m$ through the development of a thermistor model applicable to all the sensors studied so far. During the search for the causes of such a combination of effects, several hypotheses were evaluated, and many of them led either to a negligible variation of the impedances $Z_{\mathrm{th},i}$ or to the necessary appearance of nonlinearities that are not observed in the sensors. Only the dc operating point of the thermistor changes between states---bias current, dissipated power and bead temperature, related to one another through the thermal law of the element---and any dependence of the RF impedance on that operating point is compatible with the signal linearity observed in each state. Among the operating-point variables, the dc electric field is excluded: a conduction dependent on the field would be modulated by the RF field superposed on it and would appear as a nonlinearity, which is not observed; the bead temperature is the remaining variable, and the one that governs the conductivity of the material. It is constant within each state: the bridge holds it fixed, so that each state sets a new set of values for the impedance of the bead, which then remain fixed and leave the linearity of the sensor unaltered. That dependence is established below with sweeps of the bridge resistance, and a thermal law of the parallel resistance of the bead is obtained from them.

This work reports four results. First, the per-state correction $m$, computed for the eleven sensors of the preceding works, exhibits a common pattern within each family: one direction for all the specimens and a magnitude set by the family, a few tenths of a percent in the coaxial mounts and one to two percent in the waveguide mounts. Second, the corrected scattering matrix and the efficiencies of the non-central states, which the first pass leaves undetermined, are obtained, with their uncertainty. Third, sweeps of 17 to 31 states on four sensors (one of them treated in Appendix~\ref{app:sp4}) establish the properties of the correction: the corrected impedances of a sensor form one passive two-port that any triple of states recovers, the calibration factor does not depend on the triple chosen, and the dependence on the state beyond the bridge resistance is measured without any model, through the cross-ratio of four states, and is represented by the model as a change of the parallel branch of the bead. Fourth, the correction predicts the efficiency ratios between states, and these are tested on one waveguide mount by substituted-power measurements in the three states with a stable source (Section~\ref{sec:pred}): the measured ratios agree with the corrected ones within the uncertainty and exclude those of the first pass.

The results of the first pass are not repeated here, and neither are their uncertainty budgets: the uncertainty of the measured reflection coefficients, of the first-pass models and of the calibration factor is treated in full in the preceding works~\cite{greco2026coax,greco2026wg}, and the present paper builds on those results without altering them. What is budgeted here, with the techniques of the GUM~\cite{gum2008,gums1} as developed there, are the quantities this paper introduces---the correction and the efficiency ratios between states, the bead temperatures and the per-state constants of the thermistor---and the intervals are given where they are indispensable for the assessment of a result.

\section{Results of the $m$ Transformations}
\label{sec:results}
The correction~\eqref{eq:minnorm} was computed for the eleven sensors of the preceding works: the five coaxial mounts of~\cite{greco2026coax} (specimens~1, 4 and~5 of model 478A, 0.5--10~GHz; specimens~2 and~3 of model 8478B, 0.5--18~GHz) and the six waveguide mounts of~\cite{greco2026wg}: the five K486 (specimens~1--5, WR-42, 18--26.5~GHz) and the R486 (WR-28, 26.5--40~GHz). The specimen numbers are those of the respective papers, and the first-pass models, reflection coefficients and frequency grids are those reported there; nothing was re-measured or re-fitted.

In both families the states are set by the same power-meter bridge, $R_{b}=100$, 200 and 300~$\Omega$ for $i=1,2,3$, with $i=2$ the nominal operating point, and the model impedances $Z_{\mathrm{th},i}$ are evaluated at the thermistor itself. What differs is how the bridge resistance is shared. The waveguide mounts carry a single bead, which therefore operates at $R_{b}$. The coaxial mounts carry two beads in series at dc and in parallel at radio frequency, so that each bead operates at $R_{b}/2$ and the pair presents $R_{b}/4$ to the line. State~$i$ thus denotes the same bridge setting in both families, but a bead temperature and a dc resistance that are not the same: the waveguide bead at 100, 200 and 300~$\Omega$, the coaxial bead at 50, 100 and 150~$\Omega$. This distinction matters for the temperatures of Section~\ref{sec:temp}, where the state enters through the bead resistance and not through the bridge setting.

\begin{figure*}[!t]
\centering
\includegraphics[width=\textwidth]{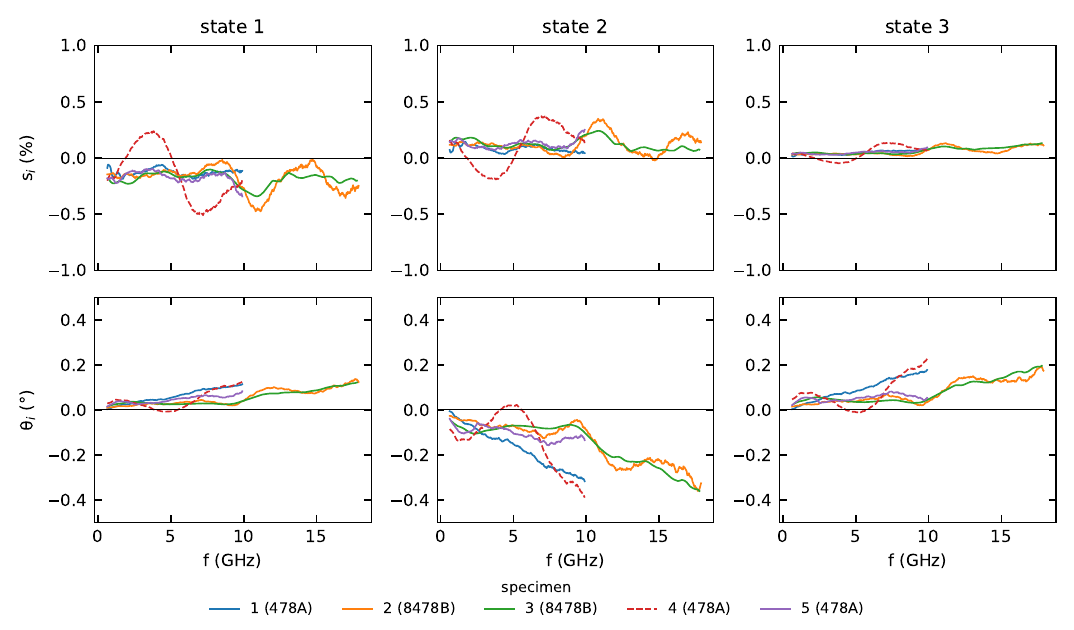}
\caption{Per-state correction of the five coaxial mounts of~\cite{greco2026coax}: logarithmic modulus $s_i$ (top) and phase $\theta_i$ (bottom) of $m_i$ against frequency, one column per state. Specimen~4 dashed. Curves are 11-point moving averages.}
\label{fig:mcoax}
\end{figure*}

\begin{figure*}[!t]
\centering
\includegraphics[width=\textwidth]{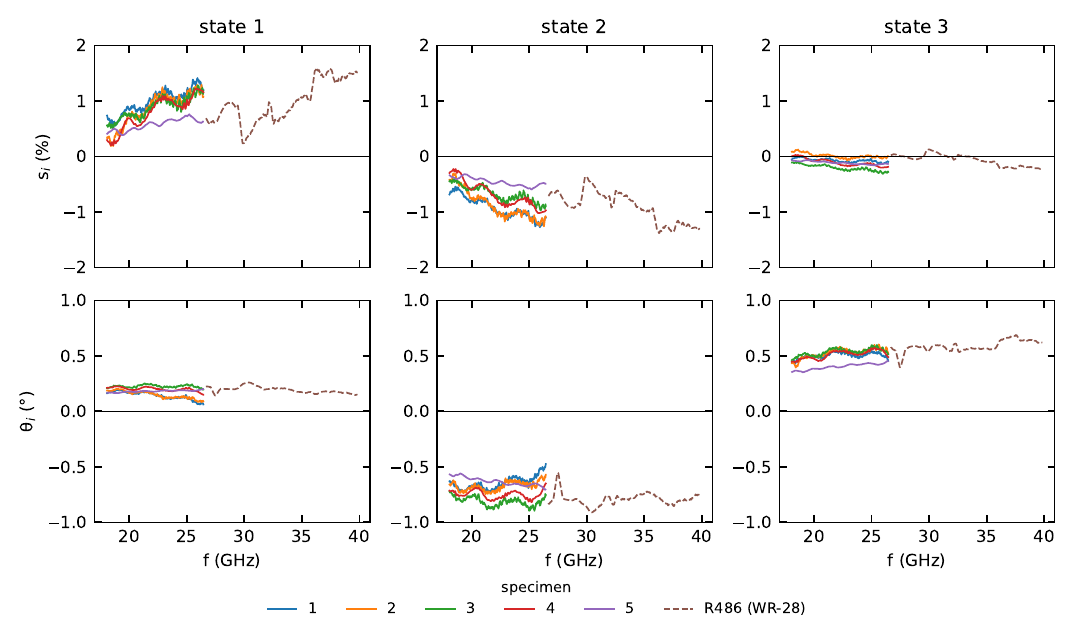}
\caption{Per-state correction of the six waveguide mounts of~\cite{greco2026wg}: specimens~1--5 (K486, WR-42) and the R486 (WR-28, dashed); same layout as Fig.~\ref{fig:mcoax}.}
\label{fig:mwg}
\end{figure*}

Table~\ref{tab:m} summarizes the result for each sensor: the spectral norm of the first-pass two-port, the band medians of the three factors $m_i$ and of $\|m-1\|=(\sum_i s_i^2+\theta_i^2)^{1/2}$, and the efficiency ratios $\eta_1/\eta_2$ and $\eta_3/\eta_2$ at the floor. Figs.~\ref{fig:mcoax} and~\ref{fig:mwg} show $s_i(f)$ and $\theta_i(f)$ for the two families, with the specimens of each superimposed; these two figures are the object of the analysis that follows. The two-port of coaxial specimen~1 in~\cite{greco2026coax} is reported after a passivity correction; Table~\ref{tab:m} gives the uncorrected first pass of the same specimen, and the selection~\eqref{eq:minnorm} is the form that correction takes here. With three states the six real unknowns of~\eqref{eq:minnorm} are fixed entirely by the two constraints and the minimum norm, so that Table~\ref{tab:m} is the correction under that convention; the efficiency ratios $\eta_1/\eta_2$ and $\eta_3/\eta_2$, which are what a calibration uses, are the same under the three-state and the $N$-state conventions to 0.03\,\%, whereas the $m_i$ are not; what the measurement itself determines when more states are added is examined in Section~\ref{sec:props}.

Three features are visible before any analysis. The first-pass two-port is active in every sensor: at its band maximum $\sigma_{\max}$ exceeds unity by 1.2--2.7\,\% in the coaxial mounts and by 12--15\,\% in the waveguide mounts (Table~\ref{tab:m}). The factors $m_i$ are small---a few tenths of a percent in the coaxial mounts, one to two percent in the waveguide mounts, which for an impedance is a small quantity by the standards of a reflection measurement; the ordinates of Figs.~\ref{fig:mcoax} and~\ref{fig:mwg} are set to $\pm1$ and $\pm2\,\%$ so that the curves are read at that scale---and at every frequency their logarithmic moduli add to zero, $\sum_i s_i=0$, a property established in Appendix~\ref{app:traceless}. And within each family the specimens, measured and modeled independently, follow one pattern: in the coaxial mounts $s_i$ is nearly constant over 0.5--10~GHz, in the proportion $s_1:s_2:s_3\approx-3:2:1$, while $\theta_i$ is linear in frequency, with slopes of opposite sign in the operating state and in the other two; in the waveguide mounts $s_1\approx-s_2$ grows with frequency, from 0.5\,\% at 18~GHz to 1.5\,\% at 40~GHz, with $s_3$ smaller by a factor of four or more, while $\theta_i$ is nearly constant, $(+0.2,-0.7,+0.5)^\circ$; the R486, with a different bead, a different waveguide and the adjacent band, continues the curves of the five K486 within the dispersion among them. The two models of coaxial mount, whose beads differ in size, follow the same pattern in Fig.~\ref{fig:mcoax} over their common band, the 478A specimens with a larger $\theta_2$ at the top of their band. The waveguide specimen~5 shows the smallest correction of its family. One coaxial specimen (specimen~4) departs from the pattern in a way examined in Appendix~\ref{app:sp4}: its $s_i$ oscillate in frequency about the common values, in the same proportion $-3:2:1$.

\begin{table*}[!t]
\caption{Per-State Correction of the Eleven Sensors: Band Medians. $\sigma_{\max}$ of the First-Pass Two-Port, Before the Correction, as Mean/Maximum over the Band (at the Floor It Equals $\sqrt{\eta_2}+\varepsilon$, $\varepsilon=10^{-3}$, Section~\ref{sec:intro}); $\|m-1\|$ in \%; $m_i$ as ($s_i$ in \%, $\theta_i$ in degrees); Efficiency Ratios at the Floor}
\label{tab:m}
\centering
\footnotesize
\setlength{\tabcolsep}{4pt}
\begin{tabular}{clccccccccc}
\toprule
Sp. & Model & Band (GHz) & $\sigma_{\max}$ (first pass) & $\|m-1\|$ (\%) & $m_1$ & $m_2$ & $m_3$ & $\eta_1/\eta_2$ & $\eta_3/\eta_2$\\
\midrule
\multicolumn{10}{l}{\textit{Coaxial mounts}~\cite{greco2026coax}}\\
1 & 478A  & 0.5--10 & 1.008 / 1.024 & 0.37 & ($-0.13$, $+0.06$) & ($+0.08$, $-0.16$) & ($+0.05$, $+0.09$) & 0.995 & 0.998\\
2 & 8478B & 0.5--18 & 1.005 / 1.027 & 0.31 & ($-0.15$, $+0.04$) & ($+0.11$, $-0.11$) & ($+0.04$, $+0.06$) & 0.995 & 0.998\\
3 & 8478B & 0.5--18 & 1.004 / 1.022 & 0.36 & ($-0.17$, $+0.04$) & ($+0.11$, $-0.10$) & ($+0.05$, $+0.05$) & 0.993 & 0.998\\
4 & 478A  & 0.5--10 & 1.000 / 1.021 & 0.34 & ($-0.18$, $+0.04$) & ($+0.13$, $-0.11$) & ($+0.04$, $+0.06$) & 0.989 & 0.999\\
5 & 478A  & 0.5--10 & 1.002 / 1.012 & 0.32 & ($-0.17$, $+0.05$) & ($+0.12$, $-0.11$) & ($+0.05$, $+0.05$) & 0.996 & 0.998\\
\midrule
\multicolumn{10}{l}{\textit{Waveguide mounts}~\cite{greco2026wg}}\\
1 & K486 & 18--26.5 & 1.086 / 1.116 & 2.00 & ($+1.00$, $+0.15$) & ($-0.93$, $-0.65$) & ($-0.07$, $+0.51$) & 0.996 & 0.998\\
2 & K486 & 18--26.5 & 1.088 / 1.133 & 2.01 & ($+0.92$, $+0.15$) & ($-0.92$, $-0.68$) & ($+0.01$, $+0.53$) & 0.989 & 1.001\\
3 & K486 & 18--26.5 & 1.101 / 1.140 & 2.10 & ($+0.86$, $+0.23$) & ($-0.66$, $-0.81$) & ($-0.20$, $+0.53$) & 0.995 & 0.999\\
4 & K486 & 18--26.5 & 1.100 / 1.123 & 2.03 & ($+0.87$, $+0.20$) & ($-0.75$, $-0.75$) & ($-0.12$, $+0.51$) & 0.996 & 0.998\\
5 & K486 & 18--26.5 & 1.088 / 1.118 & 1.55 & ($+0.59$, $+0.18$) & ($-0.47$, $-0.64$) & ($-0.12$, $+0.40$) & 0.994 & 1.000\\
6 & R486 (WR-28) & 26.5--40 & 1.098 / 1.148 & 2.15 & ($+0.93$, $+0.19$) & ($-0.88$, $-0.79$) & ($-0.05$, $+0.58$) & 0.995 & 0.996\\
\bottomrule
\end{tabular}
\end{table*}

\begin{figure*}[!t]
\centering
\includegraphics[width=\textwidth]{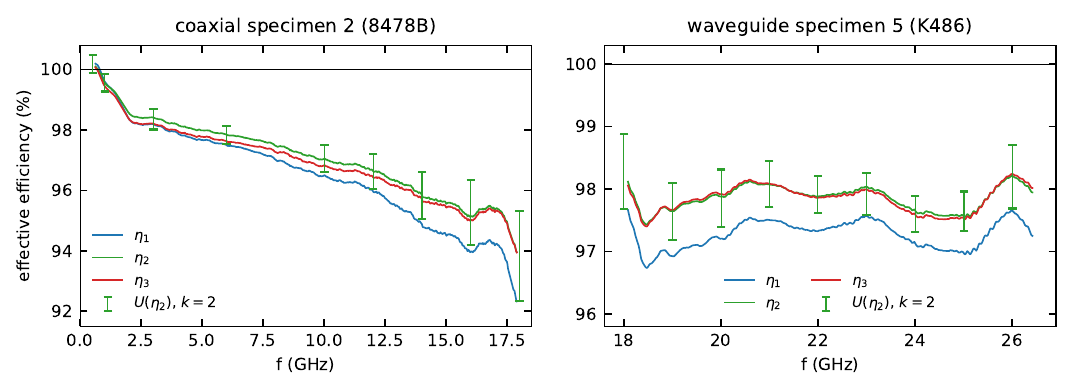}
\caption{Effective efficiencies of the three states at the passivity floor, coaxial specimen~2 of~\cite{greco2026coax} and waveguide specimen~5 of~\cite{greco2026wg}, with the expanded uncertainty ($k=2$) of $\eta_2$ from the budgets of those works, at the frequencies tabulated in~\cite{greco2026coax} for the coaxial mount and at 1~GHz intervals for the waveguide mount. Curves are 11-point moving averages.}
\label{fig:eta}
\end{figure*}
\begin{figure*}[!t]
\centering
\includegraphics[width=\textwidth]{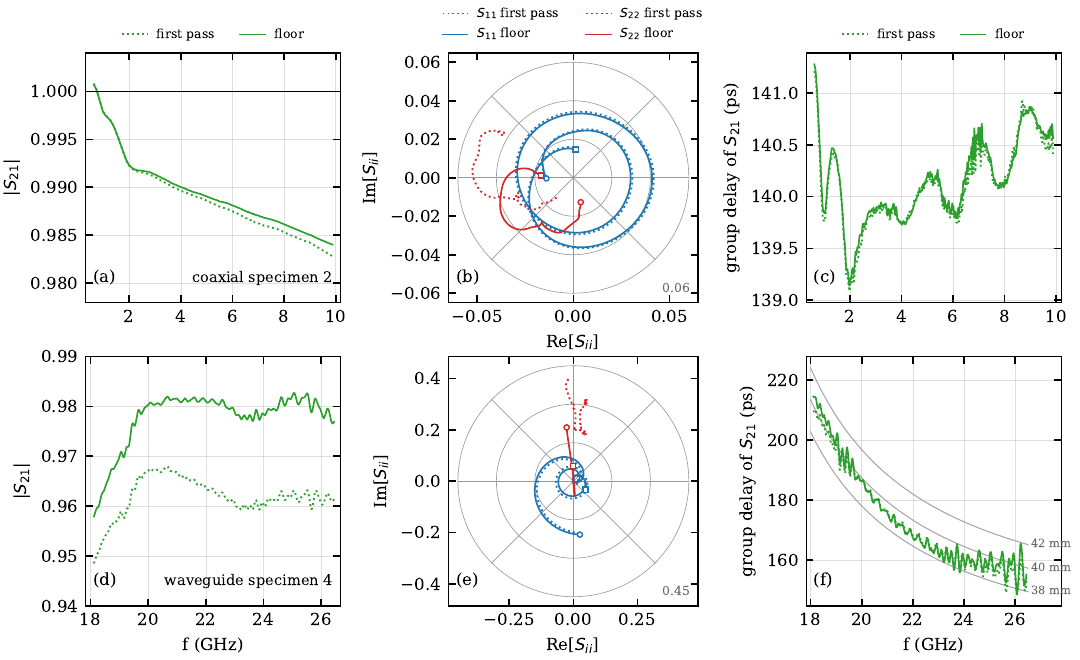}
\caption{Two-port of the first pass (dotted) and at the passivity floor (solid), three-state convention, for coaxial specimen~2 (a)--(c) and waveguide specimen~4 (d)--(f): $|S_{21}|$; $S_{11}$ and $S_{22}$ in the reflection plane (circle, start of the band; square, end; the radius of the outer circle is marked); group delay of $S_{21}$ (gray: the delay of a fixed length of WR-42, 38, 40 and 42~mm). The correction leaves the delay, the electrical length of the body, unchanged, and brings the reflection of the bead port down to that of the sensor port.}
\label{fig:Sfp}
\end{figure*}
Fig.~\ref{fig:eta} closes the presentation of the results with the effective efficiencies of the three states at the passivity floor for one specimen of each family. The three curves are smooth and lie within 1\,\% of one another over most of the band, the hot state falling 1.7\,\% below the other two at the top of the coaxial band, $\eta_2$ being that of the first pass by construction ($\eta_2$ of the first pass exceeds unity by 0.1--0.2\,\% at 0.5--0.7~GHz in the coaxial specimen, within its expanded uncertainty, and the floor follows it); the ratios $\eta_1/\eta_2$ and $\eta_3/\eta_2$ carry the repeatability of the analyzer propagated through the correction, 0.01--0.03\,\%, and the offset of the floor, up to 0.02\,\%, while the error model of the analyzer of~\cite{greco2026coax,greco2026wg} (directivity, tracking and source match as one complex term each per calibration, common to the three states), propagated through the selection, moves their band means by 0.01--0.02\,\% ($k=2$, waveguide specimen~4): an error common to the states is absorbed by the two-port; the ratios $\eta_1/\eta_2$ and $\eta_3/\eta_2$ are the quantities tested by substituted-power measurements in Section~\ref{sec:pred}. The correction therefore leaves the published results untouched and completes the description of the non-central states: the efficiencies $\eta_1$ and $\eta_3$ of the original triple and, with them, the calibration factors $K_{f,i}=\eta_i(1-|\Gamma_i|^2)$ of those states~\cite{greco2026coax}, which the first pass leaves undetermined. Fig.~\ref{fig:Sfp} shows the two-port itself for one specimen of each family, before and after the correction: $|S_{21}|$ moves by the amount that brings $\sigma_{\max}$ to the floor, the group delay of $S_{21}$, which is the electrical length of the body (140~ps in the coaxial mount; that of 40~mm of WR-42 in the waveguide mount, 160--210~ps over the band), is unchanged to 0.1 and 1~ps in band median, and the reflection of the bead port is brought down to that of the sensor port, from 0.4 to 0.2 at the lower edge of the waveguide band.

The variation of the factors $m_i$ is consistent across all the groups of sensors, and its common factor is the bead temperature: in every mount the hottest state, $T_1>T_2>T_3$, is the one whose efficiency departs most from the other two. Although a single WR-28 specimen was available, placing it next to the K486 series in Fig.~\ref{fig:mwg} shows the same trend continued, from which we infer that both are built, at least, of the same thermistor material. The temperature dependence is even clearer in the efficiencies of the three states (Fig.~\ref{fig:eta}). The transformation $m$ also carries residues of other perturbations, which we take to be of higher order; they influence the final values but are not the object of this study, and pursuing them further would not add relevant information on these sensors. It remains, however, to bound that residual error once the state dependence of the thermistor is established (Section~\ref{sec:state}).

Three bridge settings resolve the dependence on the state only at three points. To resolve it further, three sensors were measured in many states in one session each: coaxial specimens~2 and~3 and waveguide specimen~4, with $R_b$ from 100 to 400~$\Omega$ in steps of 10~$\Omega$ (31 states for specimens~3 and~4; 23 for specimen~2, which was measured at the 21 settings from 100 to 300~$\Omega$ and at 350 and 400~$\Omega$); the power-meter bridge was modified for this purpose by adding resistors, so that $R_b$ can be set at any of these values, its three nominal settings being unchanged; the resistance of every setting was measured at the mount terminals as the manufacturer's manual prescribes~\cite{hp432a}, with the resistance of the cable measured and subtracted, as in~\cite{greco2026wg}. The reflection coefficients were measured with the HP~8510C analyzer of the preceding works in its 801-point sweep, read over the GPIB bus and processed with METAS VNA Tools~\cite{vnatools} (45~MHz--18~GHz for the coaxial mounts, of which 0.5--10~GHz is used, the range of the first pass of~\cite{greco2026coax}, with the calibration of that work, an 85054D type-N kit with a sliding load; 17--26.5~GHz for the waveguide mount, of which 18--26.5~GHz is used, with the calibration of Section~\ref{sec:pred}); each sweep was completed in one session, of 30~min for the waveguide mount and of 1.5 and 3.5~h for the two coaxial mounts, with the procedure per state (bridge setting, settling and reading) of~\cite{greco2026coax,greco2026wg}. The correction was computed jointly for all the states of each sensor: the first-pass model of the sensor supplies $Z_{\mathrm{th},i}$ at every $R_b$, the two-port is required to be the same for all the states, $F$ is fixed on the nominal triple, and the floor and the minimum norm are those of~\eqref{eq:minnorm}. The additional states contribute two real constraints each, so that the $m_i$ are determined by the measurement up to six degrees of freedom, which the convention fixes. The measurement also verifies the bridge: a bilinear fit of $\Gamma(R_b)$ at each frequency reproduces the reflection coefficients of the 23 or 31 states to $2$--$4\times10^{-4}$ in the coaxial mounts and returns every $R_{b,i}$ within 0.05~$\Omega$ of its nominal value, except the two settings of specimen~3 noted in Section~\ref{sec:state}; in the waveguide mount the same fit leaves $1\times10^{-3}$ rms, the departure from bilinearity that Section~\ref{sec:state} measures, and the verification rests on the measured resistance of each setting.

\begin{figure}[!t]
\centering
\includegraphics[width=\columnwidth]{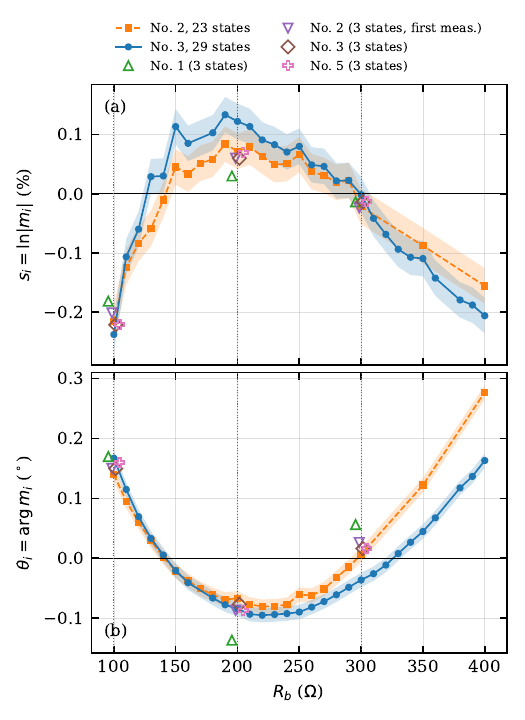}
\caption{Per-state correction of the coaxial family: band means (0.5--10~GHz) of $s_i$ (a) and $\theta_i$ (b) against the bridge resistance for specimens~2 (23 states) and~3 (29 states), with their expanded uncertainty, $k=2$ (shaded), and the three-state values of specimens~1, 2 (first measurement), 3 and~5 referred to the same degrees of freedom (open symbols).}
\label{fig:sweepcoax}
\end{figure}

\begin{figure}[!t]
\centering
\includegraphics[width=\columnwidth]{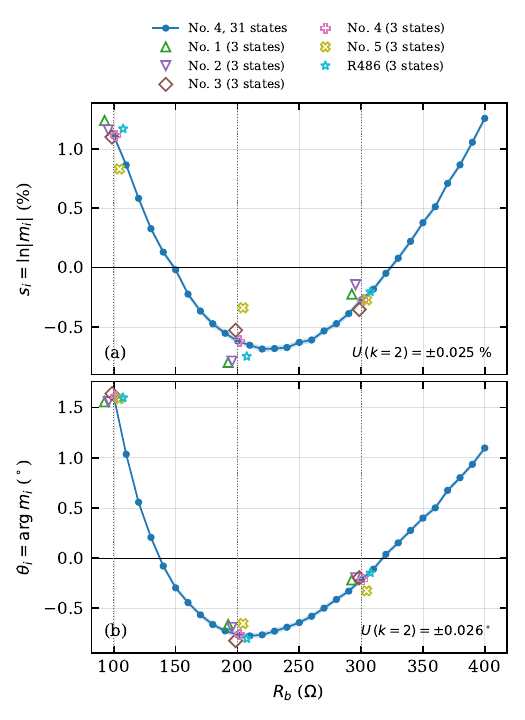}
\caption{Per-state correction of the waveguide family: band means (18--26.5~GHz) of $s_i$ (a) and $\theta_i$ (b) against the bridge resistance for specimen~4 (31 states) with its expanded uncertainty, $k=2$ (shaded band, narrower than the line; its value is written in each panel), and the three-state values of the six specimens of Table~\ref{tab:m} referred to the same degrees of freedom (open symbols).}
\label{fig:sweepwg}
\end{figure}

Figs.~\ref{fig:sweepcoax} and~\ref{fig:sweepwg} show the band means of $s_i$ and $\theta_i$ against $R_b$ for each family, together with the three-state values of the remaining specimens of Table~\ref{tab:m}, referred to the six degrees of freedom of the swept sensor: the three-state solution of each specimen is displaced, in $s_i$ and in $\theta_i$, by the difference at 100, 200 and 300~$\Omega$ between the joint solution of the swept sensor and the three-state solution of that same sensor, so that every specimen is read with the convention of the sweep (the displacement reaches $1.4^\circ$ in $\theta_1$ of the waveguide family, larger than the three-state values themselves, and the open symbols of the swept sensor lie on its own curve by construction). The shaded bands are the expanded uncertainty ($k=2$) of each mean: 0.03\,\% in $s_i$ and 0.012$^\circ$ in $\theta_i$ in the coaxial mounts, 0.025\,\% and 0.026$^\circ$ in the waveguide mount. The repeatability of the analyzer, $3$--$9\times10^{-5}$ per point in the coaxial band and $1.0$--$1.3\times10^{-4}$ in the K band, propagated by Monte Carlo through the joint solution, contributes 0.01--0.03\,\% to each $m_i(f)$ and less than 0.002\,\% to a band mean; the bands are therefore the dispersion of the means about a smooth curve in $\ln R_b$, which is what the drift of the analyzer within the session leaves. The margin $\varepsilon$ of the floor, varied between 0 (where the constraint is attainable only to the tolerance of the solver, Appendix~\ref{app:floor}) and $2\times10^{-3}$, moves the means by 0.006\,\% in the central states and by 0.02\,\% at the ends of the coaxial curves; this is a matter of convention rather than of measurement and is not included. Three facts are read from the figures. The correction is a smooth function of the state: in the coaxial mounts $s_i$ rises from $-0.23\,\%$ at 100~$\Omega$ to a maximum of $+0.08$--$0.13\,\%$ between 190 and 210~$\Omega$ and falls to $-0.16$ to $-0.21\,\%$ at 400~$\Omega$, $\theta_i$ mirrors it with the opposite sign, and both vanish near 140 and 300~$\Omega$; in the waveguide mount the same curve appears with six times the amplitude and the opposite sign of $s_i$, $+1.1\,\%$ at 100~$\Omega$, a minimum of $-0.7\,\%$ at 230~$\Omega$ and $+1.3\,\%$ at 400~$\Omega$, with zeros near 140 and 320~$\Omega$. The correction is nearly independent of frequency in the coaxial mounts over 0.5--10~GHz (standard deviation of $s_i(f)$ over the band 0.02\,\%, of $\theta_i(f)$ 0.03$^\circ$, except at the two ends of the range), while in the waveguide mount $s_i$ grows with frequency, as in~\cite{greco2026wg}, and $\theta_i$ does not. And the state at which the first pass requires the least correction lies close to, but not at, the nominal operating point: the extremum of $s_i$ is at 190--230~$\Omega$. The three bridge settings of the preceding works take one point of each region, the hot end, the extremum and a zero, which is the proportion $-3:2:1$ of Fig.~\ref{fig:mcoax}; the three-state values of the remaining specimens fall on the swept curves within 0.015\,\% in $s_i$ and $0.02^\circ$ in $\theta_i$ in the coaxial family (0.04\,\% and $0.07^\circ$ for specimen~1) and within 0.3\,\% and $0.11^\circ$ in the waveguide family, where specimen~5 has a smaller amplitude than the others. The two coaxial specimens swept differ from each other by 0.05\,\% in $s_i$ at the extremum and by up to $0.11^\circ$ in $\theta_i$ at 400~$\Omega$, the specimen-to-specimen variation within one model, of the same order as the scatter of the three-state values about the curves. Coaxial specimen~4, which departs from the others, is treated in Appendix~\ref{app:sp4}.

\subsection{Properties of the Correction}
\label{sec:props}
The sweeps allow the properties of the transformation~\eqref{eq:m}--\eqref{eq:minnorm} to be stated beyond the three states for which it was introduced; the numbers are those of coaxial specimen~2 unless stated otherwise.
\begin{enumerate}[label=(\roman*)]
\item $K_f$ and $\eta_2$ are invariant, by construction.
\item The logarithmic moduli add to zero over all the states of a solution, $\sum_i s_i=0$, to $10^{-5}$ (Appendix~\ref{app:traceless}). This is a property of the selection: a common scale of all the impedances is a change of $Z_0$ that leaves $F$ and the cross-ratios invariant and $\sigma_{\max}$ unchanged to first order, so the minimum norm assigns it no component; the first pass thus leaves the constants of $Z_{\mathrm{th}}$ at the logarithmic center of the corrected impedances, with one nonlinearity per state around it.
\item Consistency. The corrected impedances $Z'_i=m_iZ_{\mathrm{th},i}$ of the $N$ states define one two-port at the floor, and any triple of them, including three contiguous states, returns that two-port and the same efficiency of every state: $\eta(R_b)$ recomputed from the triples $(R_b-10,R_b,R_b+10)$~$\Omega$ with the $Z'_i$ agrees with the joint solution to $10^{-5}$ at the 19 states that have both neighbors at $\pm10$~$\Omega$. The first-pass impedances, not constrained to a common two-port, give efficiencies that differ between triples by the departure from bilinearity of Section~\ref{sec:state}: 0.1\,\% in the coaxial mounts, 0.3--0.9\,\% in the waveguide mount.
\item Independence of the triple. The efficiency $\eta_2$ of the operating state obtained from the corrected impedances $Z'_i$, with $F$ fixed anew on the triple chosen, for any triple of the sweep---120/200/280, 150/200/250, 170/200/230, 100/200/400, 110/200/350~$\Omega$---agrees with the nominal one within $\pm0.04\,\%$ in band mean and $\pm0.1\,\%$ per point; $K_f$ does not depend on the states chosen.
\item Continuity. The correction is a correction between states and vanishes with their separation: $\|m-1\|=0.28\,\%$ (the sweep session; 0.31\,\% in Table~\ref{tab:m}, from the first measurement), 0.04, 0.02 and 0.01\,\% for the triples $200\pm100$, $\pm50$, $\pm30$ and $\pm10$~$\Omega$, as the square of the separation.
\item Necessity. No fixed passive network between the reference plane and the bead removes the non-passivity of the first pass~\cite{greco2026wg}: such a network cascaded with the body is a change of the two-port that the first pass determines as a whole, and the non-passivity is a property of that whole; and the cross-ratio of four measured states departs from that of their bridge resistances (Section~\ref{sec:state}): the dependence on the state is in the measurement, not in the construction.
\item Genesis. The two-port originates in the first pass, as the bilinear map of the three model impedances onto the three measured reflection coefficients; the correction is the smallest displacement of the impedances that brings it to the passivity floor, and the $N$-state solution refines it by the same rule, reached by continuation from the three-state solution and independent of the starting point. What the correction changes in the two-port is confined to the bead side (Fig.~\ref{fig:Sfp}): $S_{11}$ and the transmission are preserved---in coaxial specimen~2, $|S_{11}|$ to $6\times10^{-4}$, $|S_{12}S_{21}|$ to 0.1\,\% and the group delay of $S_{21}$ to 0.1~ps out of 140~ps; in waveguide specimen~4, $|S_{11}|$ to $2\times10^{-3}$ and the delay to 1~ps out of 172~ps, band means---while $|S_{22}|$ falls from 0.037 to 0.029 in the coaxial mount and from 0.25 to 0.06 in the waveguide mount, where the non-passivity of the first pass resided. The efficiencies of the non-central states, $\eta_1/\eta_2$ and $\eta_3/\eta_2$, are the prediction of the correction, and Section~\ref{sec:pred} gives the measurement that tests it.
\end{enumerate}
The correction thus predicts the efficiencies of the states the first pass could not resolve. What follows is their test with a measurement independent of the reflection coefficients, substituted power in the three states; the reading of the correction in terms of the temperature of the bead comes after it.

\section{Test with Substituted Power}
\label{sec:pred}
The quantities that the selection~\eqref{eq:minnorm} predicts and the reflection measurement does not fix are the efficiencies of the non-central states. Two candidates are on the table for every sensor: the first pass, whose efficiencies vary monotonically from the hot to the cold state (coaxial specimen~2: 0.986 at 100~$\Omega$ to 0.966 at 400~$\Omega$; waveguide specimen~1: $\eta_1/\eta_2=0.941$, $\eta_3/\eta_2=1.023$), and the floor, whose efficiencies are nearly equal (coaxial specimen~2: 0.976--0.980 over the whole sweep; waveguide specimen~1: 0.996 and 0.998), with the operating state as the most efficient, as the floor requires (Section~\ref{sec:intro}). The difference, 1\,\% in the coaxial mounts and 2.5--6\,\% in the waveguide mounts, is measurable on the same mount without a calorimeter.

\begin{figure}[!t]
\centering
\includegraphics[width=0.8\columnwidth]{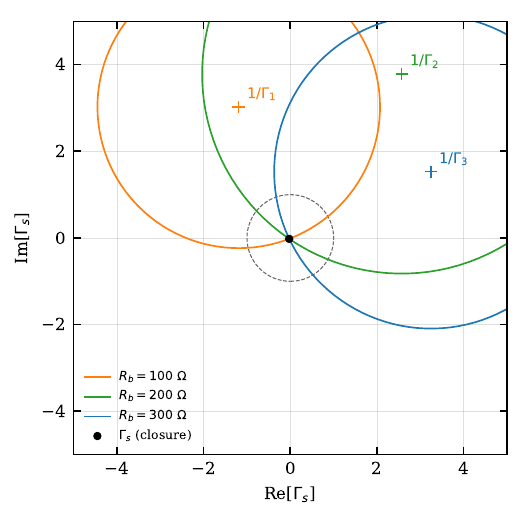}
\caption{The three substituted-power readings of waveguide specimen~1 at 22~GHz as circles in the plane of the source reflection coefficient, with centers $1/\Gamma_i$ and the scale $P_g$ of the closure; the dashed circle is $|\Gamma_s|=1$ and the dot the source reflection of the closure, $|\Gamma_s|=0.033$.}
\label{fig:apollonius}
\end{figure}

With a source of reflection coefficient $\Gamma_s$ connected at the reference plane, $P_g$ being the power it would deliver to a non-reflecting load, the power substituted in state $i$ is $P_i=P_g K_{f,i}M_i$, with $K_{f,i}=\eta_i(1-|\Gamma_i|^2)$ and $M_i=1/|1-\Gamma_s\Gamma_i|^2$; the dc power substituted is taken for the RF power absorbed by the bead, the hypothesis on which bolometric calibration and the two preceding works rest~\cite{greco2026coax,greco2026wg}, and a substitution error common to the states cancels in the ratios. The ratios between states remove the source power and the absolute calibration factor,
\begin{equation}
\frac{P_i}{P_2}=\frac{\eta_i}{\eta_2}\,\frac{1-|\Gamma_i|^2}{1-|\Gamma_2|^2}\,\frac{|1-\Gamma_s\Gamma_2|^2}{|1-\Gamma_s\Gamma_i|^2},\qquad i=1,3,
\label{eq:ratio}
\end{equation}
and leave $\Gamma_s$ as the only unknown. Each reading is a circle in the plane of $\Gamma_s$: $|1-\Gamma_s\Gamma_i|^2=P_gK_{f,i}/P_i$ is the circle of center $1/\Gamma_i$ and radius $\sqrt{P_gK_{f,i}/P_i}/|\Gamma_i|$, and the three states give three circles with one unknown scale, $P_g$ (Fig.~\ref{fig:apollonius}). The ratios~\eqref{eq:ratio} remove the scale and leave two Apollonius circles with respect to the pairs of points $(1/\Gamma_i,1/\Gamma_2)$, whose intersection is $\Gamma_s$ (of the two intersections, the one nearer the origin): at one frequency, two equations in the two components of $\Gamma_s$, which close for any pair of ratios. The test lies in requiring one physical source over the whole band, a reflection at a distance $L$ in the guide with slowly varying modulus, $\Gamma_s(f)=|\Gamma_s|(f)\,e^{\,j[\varphi_0+2L\beta_g(f)]}$, $\beta_g$ the propagation constant of the guide and $|\Gamma_s|$ quadratic in $f$; the closure residual is the distance, expressed as the relative error of the ratios, by which the circles of each frequency miss that common $\Gamma_s(f)$.

\begin{figure}[!t]
\centering
\includegraphics[width=\columnwidth]{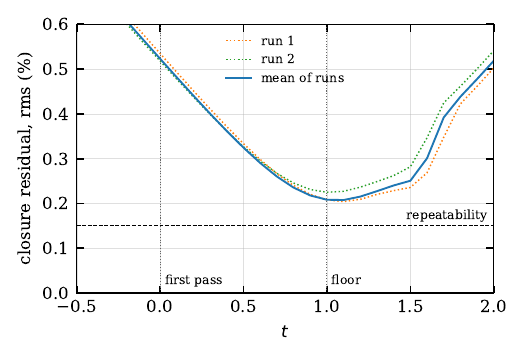}
\caption{Closure residual of~\eqref{eq:ratio} over the 18 frequencies against the position $t$ along the family joining the first pass ($t=0$) and the floor ($t=1$), waveguide specimen~1, the source reflection fitted as one reflection in the guide; the two runs and their mean; the minimum is at $t=1.06\pm0.28$ ($k=2$, Monte Carlo of the budget of Section~\ref{sec:pred}). The dashed line is the repeatability of the ratios between runs.}
\label{fig:closure}
\end{figure}

\begin{figure}[!t]
\centering
\includegraphics[width=\columnwidth]{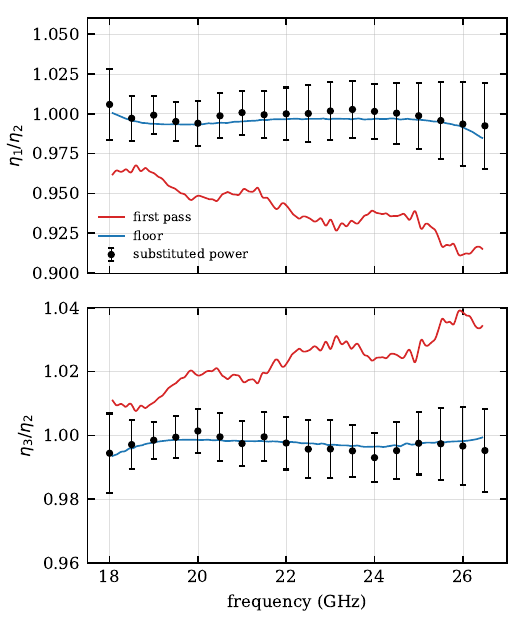}
\caption{Efficiency ratios $\eta_1/\eta_2$ (top) and $\eta_3/\eta_2$ (bottom) of waveguide specimen~1: first pass and floor from the reflection measurement (11-point moving averages), and values obtained from the substituted powers in the three states with the source reflection fixed by the closure; bars are expanded uncertainties ($k=2$).}
\label{fig:power}
\end{figure}

Waveguide specimen~1 was measured in its three states (100, 200 and 300~$\Omega$) at 18 frequencies between 18 and 26.5~GHz with a leveled synthesizer and a directional coupler as source, the mount read by its power-meter bridge (HP 432A) with an external voltmeter (Agilent 34420A) following the dc-substitution procedure of the manufacturer~\cite{hp432a}: the bridge holds the thermistor at $R_{b,i}$ in both conditions, and the RF power absorbed is the decrease of the dc power, $P_i=(V_{0,i}^2-V_{1,i}^2)/4R_{b,i}$, with $V_{0,i}$ and $V_{1,i}$ the compensation voltages without and with RF ($V_{\mathrm{comp}}$ and $V_{\mathrm{rf}}$ in the notation of the manual); the reflection coefficients measured with the HP 8510 analyzer of~\cite{greco2026wg}, in two independent runs with a repeatability of the ratios of 0.15\,\% (rms over the 36 ratios of the difference between runs, divided by $\sqrt2$). The two candidates are compared along the one-parameter family $\eta_i/\eta_2=(\eta_i/\eta_2)_{\mathrm{fp}}^{\,1-t}(\eta_i/\eta_2)_{\mathrm{floor}}^{\,t}$, in which $t$ interpolates geometrically from the first pass ($t=0$) to the floor ($t=1$) and is fitted together with the source. Along it, the closure residual of \eqref{eq:ratio} has its minimum at $t=1.06\pm0.28$ ($k=2$), with $L=3.8$~cm and a residual of 0.21\,\% (Fig.~\ref{fig:closure}); the first pass, $t=0$, leaves 0.52\,\%. The residual at the minimum remains above the repeatability of the ratios, 0.15\,\%: one reflection in the guide does not represent the source to repeatability, and a measured $\Gamma_s$ would close that gap. With $\Gamma_s$ fixed at that solution, the ratios obtained from the substituted powers at each frequency (Fig.~\ref{fig:power}) agree with those of the floor to 0.43\,\% and 0.19\,\% rms over the 18 frequencies (each ratio separately; the 0.21\,\% of the closure is the rms over the 36 relative misses of the circles) and differ from those of the first pass by 6.2\,\% and 2.7\,\%; this per-frequency comparison is a check of consistency, the test being the closure itself. The band means, $0.999\pm0.016$ and $0.997\pm0.008$ ($k=2$), contain the floor (0.996, 0.998) and exclude the first pass (0.941, 1.023). Of the two ratios, $\eta_1/\eta_2$ is the informative one: the first-pass value of $\eta_3/\eta_2$, 1.023, is already excluded by the floor, which caps every ratio at 1.002, whereas $\eta_1/\eta_2$ is left free by the floor between 0 and 1.002 and the selection places it at 0.996; the measurement confirms the cap independently and tests the selection through $\eta_1/\eta_2$. The budget comprises the statistics of the closure with $\Gamma_s$ refitted, the drift between runs, the bridge resistances, the error model of the analyzer of~\cite{greco2026wg} (directivity, source match and reflection tracking as one complex term each per calibration, common to the three states) and the phase of the flange connection ($\pm1^\circ$ per state). The analyzer was calibrated in waveguide at the flange by the offset-short, short, offset-load method, in which the reflection of the load is resolved from its two positions; the residual directivity of that calibration is then limited by the electrical length of the offset (the shim), determined with the waveguide-delay function of the analyzer to a residual phase below $0.1^\circ$, i.e.\ to 2~$\mu$m, and by the flange repeatability, and the method reaches $-60$~dB in 7~mm coaxial line at 18~GHz~\cite{rytting}. The residuals used are $-55$~dB for directivity, $-35$~dB for source match and 0.05~dB for tracking; the uncertainty program of the analyzer gives $U(\Gamma)=3.5\times10^{-3}$ ($k=2$) for $|\Gamma|\le0.4$ with this calibration, consistent with them. The statistics of the closure contribute 0.006 and 0.003 to the two ratios ($k=1$) and the directivity 0.003 and 0.002: its direct sensitivity, $2|\Gamma_i|\delta\approx10^{-3}$, is tripled because the term does not cancel in the ratios, as it nearly does in $K_f$, and is amplified by the source reflection, which is inferred from the closure rather than measured. A measured $\Gamma_s$ would remove the amplification. The bars of Fig.~\ref{fig:power} are the expanded uncertainty of each point from the same budget; the two dominant terms are common to the band, so that the bars are correlated across frequencies and the band means carry an uncertainty of the same size. A microcalorimeter of the kind used for the effective-efficiency standards~\cite{nist1511,nist1357}, operated at a second bridge setting, would give the same ratios with an uncertainty that we estimate at 0.12--0.25\,\% ($k=2$), since thermopile, line and connector cancel in the ratio.

\section{Bead Temperature in the Three States}
\label{sec:temp}
\begin{figure}[!t]
\centering
\includegraphics[width=0.9\columnwidth]{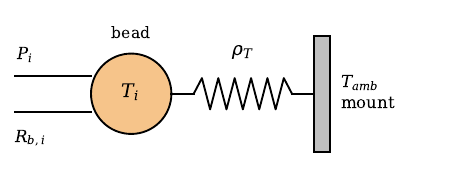}
\caption{Thermal model of the bead: bias power $P_i$ dissipated in the bead at temperature $T_i$, thermal resistance $\rho_T$ to the mount at ambient temperature $T_{amb}$; the dc resistance $R_{b,i}$ follows the law~\eqref{eq:ntc}.}
\label{fig:thermal}
\end{figure}

The state variable behind the corrections of Section~\ref{sec:results} is the temperature of the bead, which is set by the dc bias and not measured. The dc resistance of the thermistor follows the law of the NTC material,
\begin{equation}
\frac{1}{T_i}-\frac{1}{T_{amb}}=\frac{1}{B}\ln\frac{R_{b,i}}{nR_0},
\label{eq:ntc}
\end{equation}
where $T_i$ is the bead temperature in state $i$, $R_{b,i}/n$ the dc resistance of one bead---$n=2$ in the coaxial mounts, whose two beads are in series across the bridge, and $n=1$ in the waveguide mounts---, $T_{amb}$ the ambient reference temperature (the laboratory is held at $22\pm2~^\circ$C), $B$ the constant of the material and $R_0$ the resistance of one bead at the reference temperature. The bead temperature is set in turn by the dissipated power through a thermal resistance $\rho_T$ to the mount (Fig.~\ref{fig:thermal}), and that power, $P_i=V_i^2/4nR_{b,i}$, the share of one bead in the dc power of the bridge in the absence of RF, $V_i^2/4R_{b,i}$~\cite{hp432a}, follows directly from the compensation voltage $V_i$ read at the $V_{\mathrm{rf}}$ output of the meter (a BNC connector) with the voltmeter of Section~\ref{sec:pred}:
\begin{equation}
T_i=T_{amb}+\rho_T P_i .
\label{eq:selfheat}
\end{equation}
Substituting~\eqref{eq:selfheat} in~\eqref{eq:ntc} eliminates the unknown temperature,
\begin{equation}
-\frac{\rho_T P_i}{T_{amb}+\rho_T P_i}=\frac{T_{amb}}{B}\ln\frac{R_{b,i}}{nR_0},
\label{eq:comb}
\end{equation}
and each bridge setting gives one equation~\eqref{eq:comb} in the three unknowns $B$, $R_0$ and $\rho_T$, which are thus determined from the bridge voltages alone. Three settings determine them exactly, but $B$ and $\rho_T$ enter~\eqref{eq:comb} almost collinearly and three points fix their combination better than either: the sweep of Section~\ref{sec:results} removes the degeneracy. A least-squares fit of~\eqref{eq:comb} to the bridge voltages of its settings gives for the coaxial bead $B=2513$~K, $\rho_T=13\,230$~K/W and $R_0=3.04$~k$\Omega$, with a residual of 0.06\,\% in $\ln R_b$ over the range 100--400~$\Omega$ and a correlation of $-0.96$ between $B$ and $\rho_T$; the same sweep of waveguide specimen~4 (31 settings) gives $B=2671$~K, $\rho_T=10\,490$~K/W and $R_0=4.1$~k$\Omega$ with a residual of 0.08\,\%. Propagated with the correlation between $B$ and $\rho_T$ and with the $\pm2$~K of the ambient, the temperatures carry an expanded uncertainty ($k=2$) of 5--7~K in the coaxial family, where the bridge voltages were read at fourteen of the 23 settings (100, 120, 140, 150, 170, 190, 200, 220, 240, 250, 270, 290, 300 and 400~$\Omega$), and of 4--5~K in the waveguide family (31 settings), $\rho_T$ being known to 2\,\% in both; the thermal step between states, $u_i=1/T_i-1/T_2=\ln(R_{b,i}/R_{b,2})/B$, depends on $B$ alone and is known to 1\,\% in both. The fit assumes $B$ and $\rho_T$ constant over the range of the sweep, 120--300~$^\circ$C, and no radiative term; the residual of 0.06--0.08\,\% in $\ln R_b$ is the level at which the data are reproduced under them. These values serve each family and supersede the three-setting values of~\cite{greco2026coax}, which fix the combination of $B$ and $\rho_T$ rather than each. The bias powers are a property of the bridge and of the mount type and are practically the same for every specimen of a family, so one set of temperatures serves each family (Table~\ref{tab:temp}).

\begin{table}[!t]
\caption{Bias Power per Bead, Bead Temperature and Thermal State Variable $u_i=1/T_i-1/T_2$ of the Three States. Coaxial: $B=2513$~K, $\rho_T=13\,230$~K/W, $R_0=3.04$~{\normalfont k}$\Omega$ (Sweep of Specimen~2); Waveguide: $B=2671$~K, $\rho_T=10\,490$~K/W, $R_0=4.1$~{\normalfont k}$\Omega$ (Sweep of Specimen~4)}
\label{tab:temp}
\centering
\footnotesize
\setlength{\tabcolsep}{2.5pt}
\begin{tabular}{c ccc ccc}
\toprule
& \multicolumn{3}{c}{Coaxial} & \multicolumn{3}{c}{Waveguide}\\
\cmidrule(lr){2-4}\cmidrule(lr){5-7}
$i$ & $P_i$ (mW) & $T_i$ ($^\circ$C) & $u_i$ ($10^{-4}$/K) & $P_i$ (mW) & $T_i$ ($^\circ$C) & $u_i$ ($10^{-4}$/K)\\
\midrule
1 & 20.8 & 297 & $-2.76$ & 19.6 & 228 & $-2.60$\\
2 & 14.9 & 220 & 0 & 14.1 & 170 & 0\\
3 & 12.2 & 183 & $+1.61$ & 11.5 & 142 & $+1.52$\\
\bottomrule
\end{tabular}

\vspace{2pt}
\parbox{\columnwidth}{\footnotesize Expanded uncertainty ($k=2$) of $T_i$: 5--7~K in the coaxial family, 4--5~K in the waveguide family (Section~\ref{sec:temp}).}
\end{table}

Two observations follow from Table~\ref{tab:temp}. The temperatures are high and far apart: in every mount the first state is the hottest, by 60--80~K over the second, and the third the coldest, which is the order found in the corrections; the bias power per bead is nearly the same in the two families, and the coaxial bead runs some fifty degrees hotter because of the larger thermal resistance of its mounting. And the thermal step between states, $u_i=1/T_i-1/T_2=\ln(R_{b,i}/R_{b,2})/B$, is nearly the same in the two families: $u_i$ depends only on the bridge ratios and on $B$, not on $\rho_T$ or $T_{amb}$, and the two materials have close values of $B$. The circuit values of the bead are plotted below against $T_i$; the comparison between families rests on $u_i$, which the measurement fixes directly.

\section{Thermistor Parameters by State}
\label{sec:params}
\begin{figure}[!t]
\centering
\includegraphics[width=0.8\columnwidth]{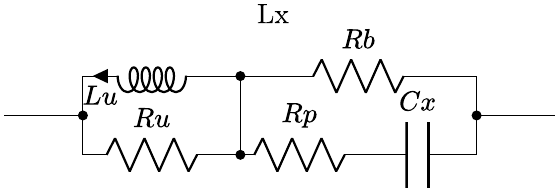}
\caption{Equivalent circuit of one thermistor, as in~\cite{greco2026coax}: dc resistance $R_b$, dielectric-loss branch $R_p$--$C_x$, and junction $R_u\,\|\,L_u$; $L_x$ is the electrical length of the bead, which sets the skin effect of $R_b$ and $R_p$, and is not an element of the drawing. $R_b$ denotes throughout the resistance set at the bridge: in the coaxial mounts each of the two beads carries $R_b/2$ at dc and the pair is in parallel at RF, so that the mount presents about $Z_{\mathrm{th}}/2$ (50~$\Omega$ at the operating state)~\cite{greco2026coax}; in the waveguide mounts the single bead carries $R_b$.}
\label{fig:circuit}
\end{figure}
The most representative model of the thermistor is the one established in each of the preceding works~\cite{greco2026coax,greco2026wg}, and the circuit is not expected to depart from it: the corrections are bounded, $|m_i-1|<2\,\%$, a small radius that the same equivalent circuit should absorb through its element values. Fig.~\ref{fig:circuit} recalls the circuit: the bead is the dc resistance $R_b$, with the skin-effect inductance $L_x$, in parallel with the dielectric-loss branch $R_p$--$C_x$, and the junction to the port is $R_u$ in parallel with $L_u$; in the waveguide model $R_p$ carries a frequency dependence $R_p(f)=R_{p0}(1-\kappa f/f_{\max})$, and the junction elements $L_u$ and $R_u$ are absent, the bead being mounted directly across the guide~\cite{greco2026wg}. We therefore keep the topology and let the elements take a value per state through constants $a_i$, $b_i$, $c_i$,
\begin{equation}
R_{p0,i}=a_i R_{p0},\qquad C_{x,i}=b_i C_x,\qquad R_{u,i}=c_i R_u ,
\label{eq:abc}
\end{equation}
where $c_i$ applies only to the coaxial mounts, whose model includes the lead resistance $R_u$; the remaining elements, among them $\kappa$ and $L_x$, keep the values of the first pass. No departure of $R_b$ with the state is considered in this section: its value is held by the bridge itself, and it is affected only by the skin effect, already treated in~\cite{greco2026coax,greco2026wg}. The constants are obtained for each state by minimizing over the band
\begin{equation}
\Phi_i=\frac{1}{N_f}\sum_{k=1}^{N_f}\frac{\left|Z'_{\mathrm{th},i}(f_k)-Z_{\mathrm{th},i}(f_k;a_i,b_i,c_i)\right|^2}{\left|Z'_{\mathrm{th},i}(f_k)\right|^2},
\label{eq:phi}
\end{equation}
so that a constant pair of values represents the correction $m_i(f)$ over the whole band, not only its band mean. At this point we formulate as a hypothesis that the per-state values are functions of the bead temperature of Section~\ref{sec:temp},
\begin{equation}
R_{p0,i}=f_r(T_i),\qquad C_{x,i}=f_c(T_i),
\label{eq:hyp}
\end{equation}
that is, that the state enters the thermistor model only through its temperature. The hypothesis rests on an elementary consideration. The bridge resistance and the bead temperature are linked one to one by the thermal law of Section~\ref{sec:temp}, so that either may label the state; but what the material of the bead is exposed to is the temperature. A dielectric, or any loss mechanism of the bead, is not sensitive to the resistance value of a neighboring element; it is sensitive to the temperature at which it sits, and the temperature is the variable that the bias sets and that reaches the dielectric.

The first minimizations showed at once that $c_i$ has no bearing on the result: in coaxial specimen~1, freeing $R_u$ together with $R_p$ and $C_x$ leaves $\sqrt{\Phi_i}$ at 0.066, 0.075 and 0.046\,\% against 0.067, 0.079 and 0.046\,\% with $R_u$ fixed, while $R_u$ and $C_x$ without $R_p$ do not reproduce the correction (0.22\,\%). The dominant values for every sensor are $R_p$ and $C_x$, that is, the dielectric-loss branch of the model. Freeing $\kappa$ as well was examined on waveguide specimen~5 and discarded: it lowers $\sqrt{\Phi_i}$ from 0.070 to 0.059\,\% in one state and not in the others, and $\kappa$ is strongly correlated with $R_{p0}$ in a fit over a finite band, the better determined quantity being $R_p$ at the band center; $a_i$ is thus reported as the ratio of $R_p$ at the band center, which with $\kappa$ fixed equals the ratio of $R_{p0}$.

Table~\ref{tab:abc} gives $a_i$ and $b_i$ for each specimen and state from the three-state corrections, referred to the first-pass values, with the residual $\sqrt{\Phi_i}$. Two constants per state represent the correction: in waveguide specimen~5 the residual falls from 0.66, 1.20 and 0.71\,\% (the departure $|Z_{\mathrm{th},i}-Z'_{\mathrm{th},i}|/|Z'_{\mathrm{th},i}|$ before the fit) to 0.07, 0.07 and 0.09\,\%, without bias in either $\mathrm{Re}[Z]$ or $\mathrm{Im}[Z]$ and without a residual frequency law. Every sensor shows the same form in $a_i$: the operating state departs downwards from the first pass, $a_2=0.92$--$0.94$ in the waveguide mounts and $0.95$--$0.96$ in the coaxial ones, while the hot and the cold states lie above it, $a_1=1.01$--$1.09$ and $a_3=1.01$--$1.04$, in the ten sensors and in the two families; the meaning of this form is examined in Section~\ref{sec:state}.

The swept sensors of Section~\ref{sec:results} settle the part of $C_x$. Fitted state by state, $b_i=1.000\pm0.003$ between 150 and 350~$\Omega$ in the coaxial mounts and $1.00\pm0.01$ between 110 and 400~$\Omega$ in the waveguide mount, and $b_i$ is the direction least constrained by the floor: moving the triple on which $F$ is fixed from 100/200/300 to 150/250/350~$\Omega$ changes $b_i$ by up to 15\,\% and $a_i$ by less than 0.01, while the two directions are orthogonal within each fit. A temperature law for $C_x$ of the Curie type, $C_x\propto1/T$, which would make $b_i=T_2/T_i$ (0.86 at 100~$\Omega$ and 1.14 at 400~$\Omega$ in the coaxial beads), is excluded by a quantity that no convention touches, the cross-ratio of four measured states (Section~\ref{sec:state}): with those $b_i$ imposed and the $a_i$ refitted, the imaginary part of the departure predicted for the wide quadruple of coaxial specimen~2 rises to $5.5\times10^{-3}$ at 9.5~GHz against $4.6\times10^{-3}$ measured ($4.3\times10^{-3}$ with $a_i$ alone), and the rms disagreement over the four quadruples of Section~\ref{sec:state} and the band rises from $0.7$ to $1.0\times10^{-3}$. $C_x$ is therefore held at its first-pass value, and the state enters the thermistor model through one constant, $a_i$; with $C_x$ fixed the $a_i$ of the swept sensors change by less than $4\times10^{-4}$ and the residual of the one-parameter fit is $0.01$--$0.22\,\%$ in the coaxial mounts and $0.06$--$0.34\,\%$ in the waveguide mount. Coaxial specimen~4, which shows a behavior of its own, is treated in Appendix~\ref{app:sp4}.

\begin{figure}[!t]
\centering
\includegraphics[width=\columnwidth]{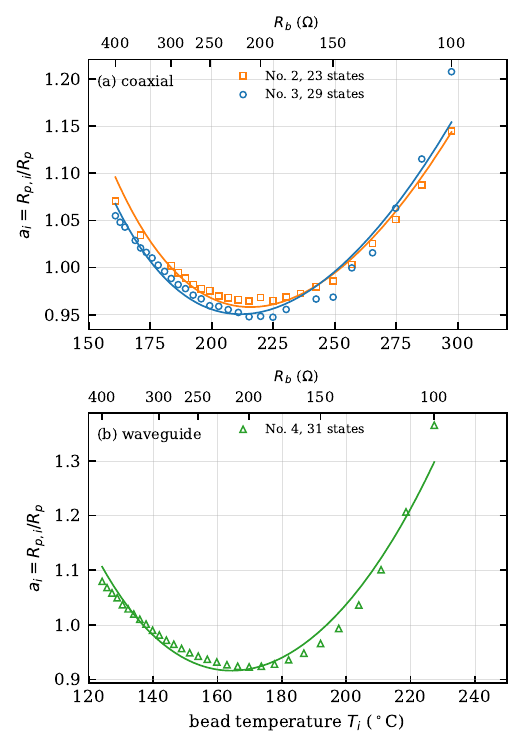}
\caption{Parallel resistance by state, $a_i=R_{p,i}/R_p$, against bead temperature for the swept sensors: coaxial specimens~2 and~3 (a) and waveguide specimen~4 (b), with the law~\eqref{eq:rplaw} fitted with one $\beta$ per family (Table~\ref{tab:rplaw}); the upper scale gives the bridge resistance $R_b$ of each state through the thermal law of Section~\ref{sec:temp}.}
\label{fig:rplaw}
\end{figure}

The three-state constants of Table~\ref{tab:abc} and the swept constants of Fig.~\ref{fig:rplaw} are not the same quantity at the hot end, and the difference is not the second constant: with $C_x$ held at its first-pass value, the three-state $a_i$ of Table~\ref{tab:abc} change by less than 0.002, the residual rising to 0.08--0.22\,\% in the coaxial mounts and to 0.1--1.0\,\% in the waveguide mounts, whereas the sweep gives at 100~$\Omega$ $a_1=1.14$ in coaxial specimen~2 and 1.36 in waveguide specimen~4, against 1.03 and 1.06 from the three states of the same sensors. The difference is the convention: with three states the six degrees of freedom of the selection~\eqref{eq:minnorm} are fixed by those three states, with $N$ states by all of them, and $a_i$ inherits that choice through $m_i$. Referred to the convention of the sweep, as the $m_i$ of Figs.~\ref{fig:sweepcoax} and~\ref{fig:sweepwg}, the three-state values of the remaining specimens become $a_1=1.13$--$1.15$, $a_2=0.957$--$0.972$ and $a_3=1.004$--$1.010$ in the coaxial family and $1.31$--$1.39$, $0.923$--$0.945$ and $0.967$--$0.987$ in the waveguide family, the values of the swept sensors at the same states. The law~\eqref{eq:rplaw} is fitted to the swept constants, which carry the convention of the whole sweep.

\begin{table}[!t]
\caption{Per-State Constants of the Dielectric-Loss Branch, Referred to the First-Pass Values, and Residual $\sqrt{\Phi_i}$ of the Fit}
\label{tab:abc}
\centering
\footnotesize
\setlength{\tabcolsep}{1.8pt}
\begin{tabular}{lcccccccc}
\toprule
Specimen & $a_1$ & $a_2$ & $a_3$ & $b_1$ & $b_2$ & $b_3$ & $\sqrt{\Phi_{1,2,3}}$ (\%)\\
\midrule
Coaxial 1 & 1.030 & 0.954 & 1.019 & 1.048 & 0.979 & 1.000 & 0.07 / 0.08 / 0.05\\
Coaxial 2 & 1.028 & 0.961 & 1.016 & 1.047 & 0.984 & 0.996 & 0.13 / 0.14 / 0.06\\
Coaxial 3 & 1.048 & 0.951 & 1.015 & 1.019 & 1.002 & 0.993 & 0.10 / 0.11 / 0.06\\
Coaxial 5 & 1.037 & 0.963 & 1.013 & 1.075 & 0.977 & 0.992 & 0.11 / 0.12 / 0.04\\
Waveguide 1 & 1.019 & 0.942 & 1.032 & 0.923 & 1.037 & 1.009 & 0.24 / 0.29 / 0.19\\
Waveguide 2 & 1.009 & 0.944 & 1.035 & 0.941 & 1.035 & 1.002 & 0.26 / 0.25 / 0.09\\
Waveguide 3 & 1.089 & 0.928 & 1.016 & 0.929 & 0.945 & 1.081 & 0.36 / 0.32 / 0.39\\
Waveguide 4 & 1.061 & 0.923 & 1.030 & 0.917 & 1.012 & 1.027 & 0.23 / 0.17 / 0.11\\
Waveguide 5 & 1.077 & 0.931 & 1.019 & 0.943 & 0.957 & 1.058 & 0.07 / 0.07 / 0.09\\
Waveguide R486 & 1.037 & 0.929 & 1.036 & 0.923 & 1.032 & 1.011 & 0.37 / 0.23 / 0.25\\
\bottomrule
\end{tabular}
\end{table}

The thermal law of $R_p$ follows from two mechanisms. The body of the bead is the NTC semiconductor whose dc resistance the bridge fixes at $R_{b,i}$: a term proportional to $R_{b,i}$ is what a fraction of that material coupled capacitively to the current path would contribute, the picture Montgomery already drew of the thermistor at microwave frequencies~\cite[Sec.~3-19, p.~149]{montgomery1947}; the sintered material is the bulk of the bead, and the hypothesis is stated as such. In series with it, the resistance of the contacts and of the remaining loss paths of the bead material grows with temperature. The parallel resistance of the model is therefore written
\begin{equation}
R_{p,i}=K\,R_{b,i}+R_c\,e^{\beta\,(T_i-T_2)},
\label{eq:rplaw}
\end{equation}
with $K$ the ratio of the RF to the dc resistance of the body, $R_c$ the positive-coefficient term at the operating temperature and $\beta$ its coefficient; referred to the first-pass value, $a_i=c\,[\rho_i+r\,e^{\beta(T_i-T_2)}]$ with $\rho_i=R_{b,i}/R_{b,2}$, $c=KR_{b,2}/R_p$ and $r=R_c/(KR_{b,2})$. The temperatures are those of Section~\ref{sec:temp}. Fig.~\ref{fig:rplaw} shows the $a_i$ of the three swept sensors against $T_i$ with~\eqref{eq:rplaw} fitted with one $\beta$ per family, and Table~\ref{tab:rplaw} gives the constants. The law reproduces the minimum near the operating state and the rise toward both ends, with residuals of 0.8 and 1.2\,\% rms in the coaxial sweeps and 1.7\,\% in the waveguide sweep, largest at the hot end, where the measured rise is steeper than exponential. Since $T_i-T_2=\rho_T(P_i-P_2)$, what the data determine is the exponent per unit of dissipated power, $\beta\rho_T=73\pm5$~W$^{-1}$ in the coaxial beads and $142\pm12$~W$^{-1}$ in the waveguide bead ($k=2$, from the covariance of the fit; $c$ and $r$ are determined to 5--10\,\%); $\beta$ itself, $(5.5\pm0.4)\times10^{-3}$ and $(1.35\pm0.12)\times10^{-2}$~K$^{-1}$, carries in addition the 2\,\% uncertainty of $\rho_T$ of Section~\ref{sec:temp}; the body accounts for 31--33\,\% of $R_p$ at the operating state in the coaxial beads and 45\,\% in the waveguide bead.

The form of $a_i$---a minimum near the operating state and a rise toward both ends---is not that of a substitution error. The mechanisms of the dc--RF substitution error of a bolometer~\cite[Sec.~6.5.1, p.~88]{fantom1990}, \cite{engen1964}---the difference between the distributions of the dc and the RF currents in the element, and between the two elements of a pair---change smoothly with the bias and give no reason for an extremum at the operating state: such an error would appear as a trend of the efficiency with the state. The first pass does show a trend of that kind (Section~\ref{sec:pred}), and it is the one the substituted-power test excludes. The minimum at the operating state is a property of the sensor as designed: the bias point recommended by the manufacturer is the one at which the parallel resistance of the bead is least, and the law~\eqref{eq:rplaw} describes how it rises on either side of it.

\begin{table}[!t]
\caption{Constants of the Law~\eqref{eq:rplaw} for the Swept Sensors (23, 29 and 31 States), One $\beta$ per Family}
\label{tab:rplaw}
\centering
\footnotesize
\setlength{\tabcolsep}{3pt}
\begin{tabular}{lcccc}
\toprule
Specimen & $c$ & $r$ & $\beta$ (K$^{-1}$) & rms (\%)\\
\midrule
Coaxial 2 & $0.316\pm0.017$ & $2.04\pm0.18$ & $5.51\times10^{-3}$ & 0.75\\
Coaxial 3 & $0.297\pm0.015$ & $2.21\pm0.18$ & $5.51\times10^{-3}$ & 1.23\\
Waveguide 4 & $0.418\pm0.020$ & $1.20\pm0.12$ & $1.35\times10^{-2}$ & 1.73\\
\bottomrule
\end{tabular}
\vspace{2pt}
\parbox{\columnwidth}{\footnotesize $c$ and $r$ with their expanded uncertainty ($k=2$); $\beta$: see text.}
\end{table}

\section{What the Sweep Determines About the Bead}
\label{sec:state}
The sweeps allow a test that involves no convention. The two-port of a sensor is fixed exactly by the nominal triple, with the first-pass model impedances and no correction, and the input reflection coefficient of every other state is predicted from its model impedance and compared with the measurement; the residual is expressed as the equivalent departure of the load, $|\delta\ln z_i|=|\delta\Gamma_i|/|\partial\Gamma/\partial\ln z|$, in band rms. In the coaxial mounts the residual is 0.01--0.05\,\% for every state between 110 and 290~$\Omega$, 0.10\,\% at 350~$\Omega$ and 0.17--0.27\,\% at 400~$\Omega$; in the waveguide mount it is 0.24--0.33\,\% between 110 and 150~$\Omega$, 0.04--0.14\,\% between 210 and 290~$\Omega$, and grows to 0.9\,\% at 400~$\Omega$. The result does not depend on the thermistor model: any load impedance proportional to $R_b$ gives the same residuals to 0.01\,\%. It is a property of the sensor seen from its port: the reflection coefficient is a bilinear function of the bridge resistance to 0.05\,\% in the coaxial mounts and to 0.1--0.3\,\% in the waveguide mount within the range of the triple, and these residuals bound the model error for any state of the bridge. This is why three states suffice, and why $\eta_2$ does not depend on which three (Section~\ref{sec:props}). The efficiency of the operating state is thus a property of the body, the two-port, and not of the thermistor model used to terminate it; the corrected two-port has the same property, and exactly: with the corrected impedances any triple recovers it and every efficiency to $10^{-5}$ (property (iii) of Section~\ref{sec:props}). What the bead itself adds is the residual, which the rest of this section measures.

\begin{figure*}[!t]
\centering
\includegraphics[width=\textwidth]{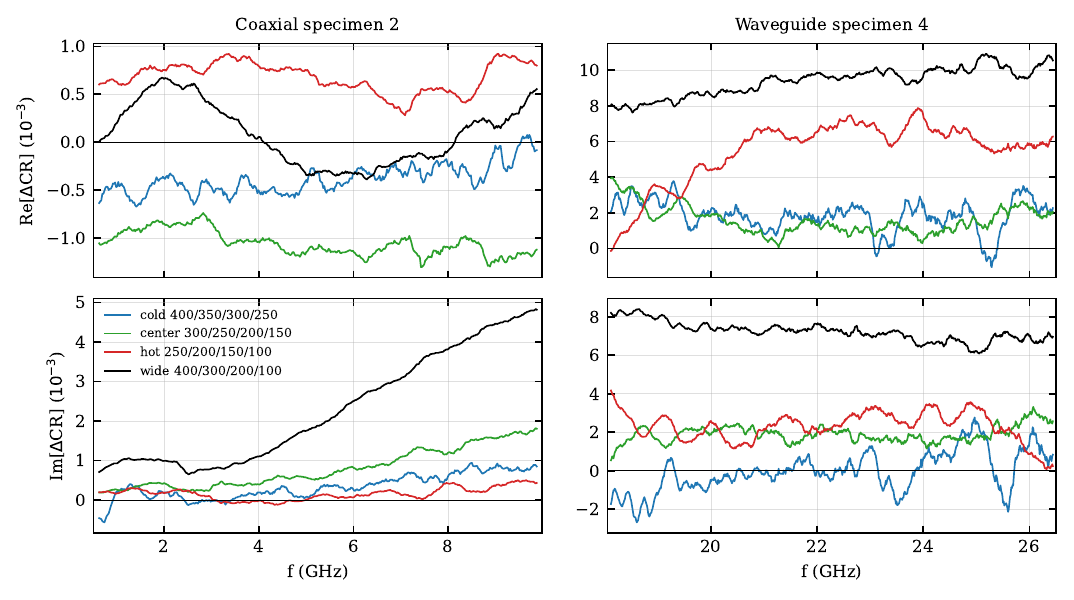}
\caption{Cross-ratio of four measured states against that of their bridge resistances, $\Delta\mathrm{CR}=\mathrm{CR}(\Gamma)-\mathrm{CR}(R_b)$, for cold, center, hot and wide quadruples of coaxial specimen~2 (left) and waveguide specimen~4 (right); real part (top) and imaginary part (bottom). Curves are 11-point moving averages.}
\label{fig:cr4}
\end{figure*}

The departure itself is measured without any model by the cross-ratio of four states, which every bilinear map leaves invariant: were the sensor bilinear in $R_b$, $\mathrm{CR}(\Gamma_1,\Gamma_2;\Gamma_3,\Gamma_4)=(\Gamma_1-\Gamma_3)(\Gamma_2-\Gamma_4)/[(\Gamma_1-\Gamma_4)(\Gamma_2-\Gamma_3)]$ would equal the real number $\mathrm{CR}(R_{b,1},R_{b,2};R_{b,3},R_{b,4})$, whatever network stands between the reference plane and the bead; a departure means that no state-independent two-port terminated in the bridge resistance reproduces the four reflection coefficients, that is, that something in the bead other than its dc resistance changes with the state. Fig.~\ref{fig:cr4} shows $\Delta\mathrm{CR}=\mathrm{CR}(\Gamma)-\mathrm{CR}(R_b)$ against frequency for quadruples of coaxial specimen~2 and waveguide specimen~4. In the coaxial mount the real part shows no trend with frequency and stays within $\pm1.3\times10^{-3}$, the error of the bridge resistors ($\pm0.05$~$\Omega$) which the sweep measures (two settings of specimen~3, 170 and 370~$\Omega$, are off by $+0.43$ and $+0.54$~$\Omega$ and are omitted from the per-state figures and fits, which use the remaining 29 states); the imaginary part is positive above 1~GHz in every quadruple and grows nearly linearly with frequency---from 0.9 to $4.8\times10^{-3}$ between 1 and 10~GHz for 400/300/200/100~$\Omega$, against a noise of $3\times10^{-4}$ per point---and faster than linearly with the span of the quadruple, as the second differences of a smooth function do; for the quadruples 50~$\Omega$ apart it reaches $0.4$--$1.8\times10^{-3}$ at 10~GHz. In the waveguide mount, with the measured bridge resistances, both parts are an order of magnitude larger and nearly flat in frequency: $8$--$11\times10^{-3}$ in the real part and $6$--$8\times10^{-3}$ in the imaginary part for the wide quadruple, larger on the hot side (up to $8\times10^{-3}$ and $4\times10^{-3}$ for 250/200/150/100~$\Omega$) than on the cold side. The per-state constants of Section~\ref{sec:params} reproduce these departures: in the coaxial mount, $\mathrm{CR}$ of the model impedances with $a_i$ alone gives $4.5\times10^{-3}$ against $4.8\times10^{-3}$ measured at 10~GHz for the wide quadruple and $1.0$ against $0.9\times10^{-3}$ for the cold one, and nothing for the hot one; in the waveguide mount the two constants $(a_i,b_i)$ give $9.2$ and $7.2\times10^{-3}$ at 22~GHz against $9.9$ and $7.6\times10^{-3}$ measured for the wide quadruple, and $5.4$ and $2.3\times10^{-3}$ against $6.5$ and $2.1\times10^{-3}$ for the hot one, the $b_i$ being those of the joint solution, within 1\,\% of unity (Section~\ref{sec:params}), while $a_i$ alone leaves the real part unaccounted for: in the waveguide bead the 1\,\% of $C_x$ carries the real part of the departure. A fixed network with the first-pass constants gives zero in every case, as it must. What the bead does beyond scaling with $R_b$ is thus a change of its parallel branch with the state, resistive and reactive in the waveguide bead and mainly reactive, growing with frequency, in the coaxial bead.

\begin{figure}[!t]
\centering
\includegraphics[width=\columnwidth]{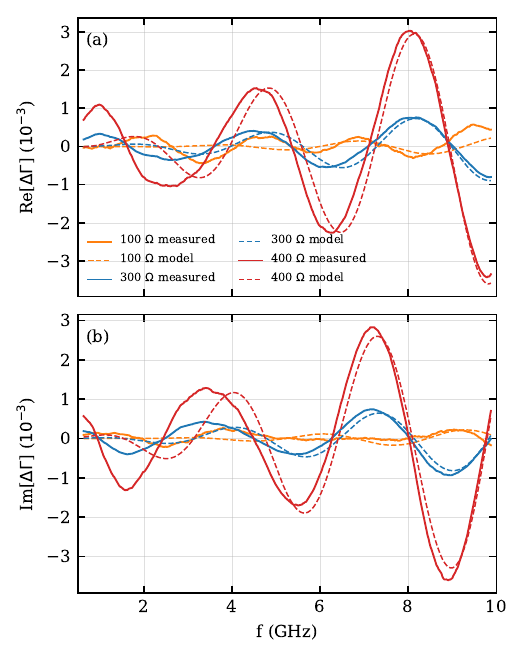}
\caption{Departure $\Delta\Gamma$ of the measured reflection coefficients of coaxial specimen~2 from the bilinear map in $R_b$ fitted on the states 150--250~$\Omega$, real (a) and imaginary (b) parts against frequency for the states 100, 300 and 400~$\Omega$ (solid, 11-point moving averages), and the same departure of the reflection coefficients predicted by the one-constant-per-state model (dashed).}
\label{fig:window}
\end{figure}

The cross-ratio compares four states; the sweep allows the same test over all of them, resolved in frequency. At each frequency a bilinear map $\Gamma=(aR_b+b)/(cR_b+1)$ is fitted by least squares to the eleven states between 150 and 250~$\Omega$, where the bead temperature changes least, and extrapolated to the other states; the departure $\Delta\Gamma_i=\Gamma_i-M(R_{b,i})$ of the measured reflection coefficients from that map is the state dependence accumulated from the window to each state. It is obtained without any model of the bead or of the two-port, from the measured $\Gamma_i$ and $R_{b,i}$ alone, at the reference plane and before any correction $m_i$ is applied, so that it does not depend on the selection~\eqref{eq:minnorm}; and it vanishes for any state-independent network terminated in $R_b$. The same map is fitted to the reflection coefficients that the per-state model predicts, $\Gamma_i^{\mathrm{mod}}=S(Z_{\mathrm{th}}(R_{b,i};a_i))$, and the two departures are compared state by state (Fig.~\ref{fig:window}). In coaxial specimen~2, $\Delta\Gamma$ is a phasor that grows with frequency and rotates with the phase of the two-port, reaching $3.5\times10^{-3}$ at 400~$\Omega$ and 10~GHz against a repeatability of $10^{-4}$ per point, and the one-constant-per-state model reproduces it in magnitude and in phase (rms $2.1$ against $2.2\times10^{-3}$ over the band, correlation 0.93 in both parts), while the first-pass model gives zero.

\begin{figure}[!t]
\centering
\includegraphics[width=\columnwidth]{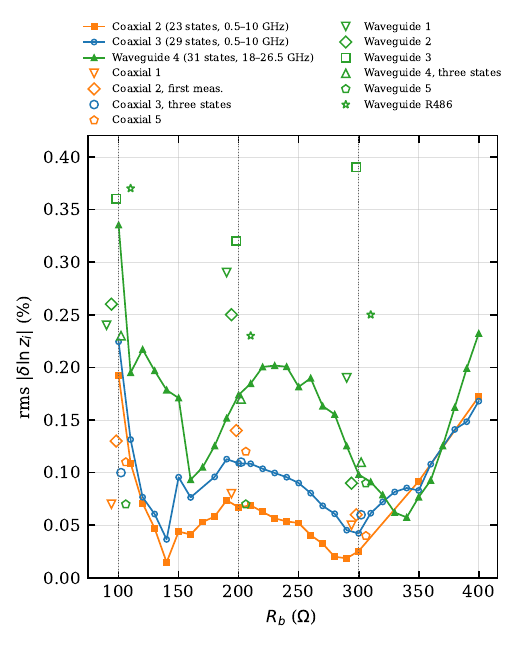}
\caption{Error of the per-state thermistor model against the measured reflection coefficients, after the transformation: band rms of $|\delta\ln z_i|$. Curves: swept sensors, one constant $a_i$ per state with $C_x$ fixed, coaxial specimens~2 and~3 (0.5--10~GHz) and waveguide specimen~4 (18--26.5~GHz). Open symbols: all sensors in their three states, with the two constants of Table~\ref{tab:abc} (displaced by a few ohms for legibility).}
\label{fig:resid}
\end{figure}

The per-state model of Section~\ref{sec:params}, one constant $a_i$ with $C_x$ fixed, reproduces this departure and interpolates it. Fig.~\ref{fig:resid} gives its residual against the measured reflection coefficients after the transformation, that is, the part of $m_i(f)$ that one constant per state does not represent: 0.01--0.19\,\% in coaxial specimen~2, 0.04--0.22\,\% in specimen~3 and 0.06--0.34\,\% in the waveguide mount. The remaining sensors are placed on the same figure with the residuals of Table~\ref{tab:abc}: with three states the selection puts part of the correction along $C_x$ (Section~\ref{sec:params}), so that their description needs the second constant, and with it they fall within 0.04--0.14\,\% in the coaxial family and 0.07--0.39\,\% in the waveguide family, the range of the swept sensors. Read against the temperatures of Section~\ref{sec:temp}, the residual that one constant per state leaves is a reactive change of the impedance growing with frequency.

\begin{figure}[!t]
\centering
\includegraphics[width=0.9\columnwidth]{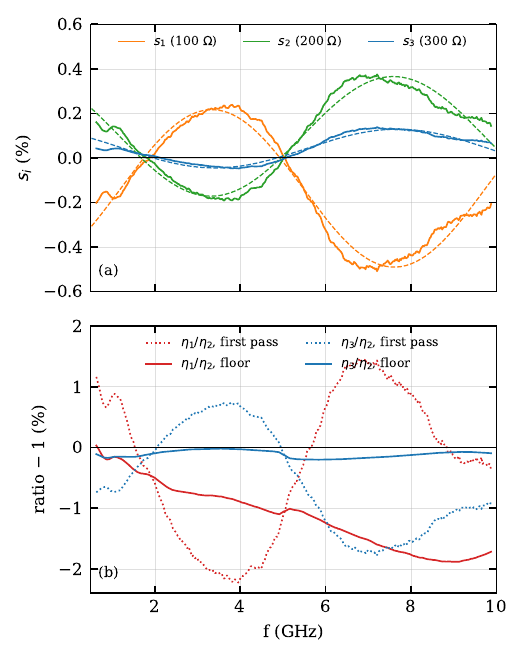}
\caption{Coaxial specimen~4, three states. (a) Logarithmic moduli $s_i(f)$ of the correction (11-point moving averages) and the approximation~\eqref{eq:sp4} (dashed). (b) Efficiency ratios of the first pass (dotted) and at the floor (solid).}
\label{fig:sp4}
\end{figure}
\begin{figure*}[!t]
\centering
\includegraphics[width=\textwidth]{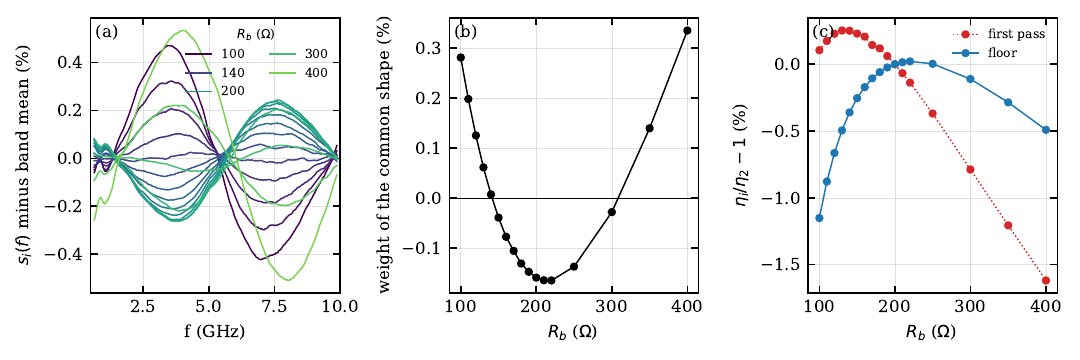}
\caption{Coaxial specimen~4, sweep of 17 states. (a) $s_i(f)$ about its band mean for every state (11-point moving averages). (b) Weight of the common shape in each state. (c) Band means of $\eta_i/\eta_2$ of the first pass and at the floor against the bridge resistance.}
\label{fig:sp4sweep}
\end{figure*}
\section{Conclusion}
The properties of the per-state transformation $m$ introduced in~\cite{greco2026wg} have been developed on the eleven sensors of the preceding works. On every sensor the correction is small, a few tenths of a percent in the coaxial mounts and one to two percent in the waveguide mounts; within each family it has one form, and the hottest state is the one whose efficiency departs most. It leaves the first-pass results of~\cite{greco2026coax} and~\cite{greco2026wg} untouched and completes them with the effective efficiencies of the non-central states, which lie within 1.1\,\% of the central one in band median, in both families.

An independent measurement supports these efficiencies. The substituted powers of the three states, with the source reflection obtained from the closure of the three Apollonius circles, give efficiency ratios that contain the values at the passivity floor and exclude those of the first pass by more than three times the expanded uncertainty: the ratios corrected by the method are the correct ones, to the level of uncertainty reached here. The test does not separate the floor from equal efficiencies, which differ from it by less than 0.5\,\%; a microcalorimeter operated at a second bridge setting would.

The sweeps in bridge resistance resolve the dependence on the state and show where it resides. The bridge fixes the dc resistance of the bead; what changes with the state is the branch the bridge does not control, the dielectric-loss branch $R_p$--$C_x$, and the sweeps show that of its two elements the reactive one does not change with the state in the coaxial beads ($b_i=1.000\pm0.003$) and changes by at most 1\,\% in the waveguide bead, so that one constant per state, the parallel resistance $R_p$, represents the correction to 0.01--0.22\,\% in the coaxial mounts and 0.06--0.34\,\% in the waveguide mount, where the 1\,\% of $C_x$ carries the real part of the cross-ratio departure. Its law~\eqref{eq:rplaw} is the superposition of two properties: a part proportional to the bridge resistance, consistent with a fraction of the NTC material coupled capacitively~\cite{montgomery1947}, and a part with a positive temperature coefficient, the rest of the body of the bead; the two appear in every sensor with one thermal coefficient per family, which points to a property of the thermistor material itself. The same dependence is what makes the cross-ratio of four measured reflection coefficients depart from that of the four bridge resistances, and the per-state constants reproduce that departure. What one constant per state leaves is a reactive change of the impedance growing with frequency, which bounds the description given here.

One sensor was analyzed apart, because it shows a behavior of its own; that behavior is documented in Appendix~\ref{app:sp4} without pursuing its physics further, since the ten others follow one pattern. Its departure did not affect the quantities that matter for calibration, the calibration factor and the efficiencies of the states, which the method obtains for it as for the others.

The measured reflection coefficients and bridge voltages of the sweeps are available from the author on request.

\appendices
\section{A Mount with a Behavior of Its Own}
\label{app:sp4}
Coaxial specimen~4, a 478A mount, is the one sensor of the eleven whose correction does not follow the pattern of its family, and this appendix gives what was measured on it. Its first-pass two-port is active, $\sigma_{\max}$ up to 1.021 (Table~\ref{tab:m}), and the correction~\eqref{eq:minnorm} brings it to the floor as in every other sensor: the efficiency ratios of the first pass oscillate between $-2.2$ and $+1.2$\,\% and between $-1.8$ and $+0.7$\,\%, in antiphase with zero crossings fixed in frequency (1.8 and 5~GHz), and at the floor they are smooth, $\eta_1/\eta_2=0.989$ and $\eta_3/\eta_2=0.999$, as in the other coaxial mounts (Fig.~\ref{fig:sp4}(b)).

What distinguishes the specimen is the form of its correction (Fig.~\ref{fig:sp4}(a)). Over 0.5--10~GHz the moduli are described by
\begin{equation}
s_i(f)=c_i+A_i\cos\!\left(2\pi f/\Delta f+\varphi_i\right),
\label{eq:sp4}
\end{equation}
with $\Delta f=8.55$~GHz common to the three states, $c_i=-0.14$, $+0.10$, $+0.04$\,\%, $A_i=-0.35$, $+0.27$, $+0.09$\,\%, the three phases equal within $5^\circ$ ($36$--$41^\circ$), and residuals of 0.02--0.06\,\%, the level of the other coaxial mounts; the phases $\theta_i$ are linear in frequency with the slopes of specimen~1. The constants $c_i$ are the correction of the coaxial family (Section~\ref{sec:results}; the band medians of Table~\ref{tab:m}, which include the modulation, differ from them by up to 0.04\,\%); the amplitudes stand in the same proportion, $-3:2:1$, and are 2.6 times larger. A second, independent measurement of the specimen, with its own calibration of the analyzer, gives the same $\Delta f$ (8.40~GHz), the same $c_i$ and $A_i$ within 0.02\,\% and the same phases within $5^\circ$; the feature is therefore of the sensor and not of the calibration. Nor does the evidence point to a residue of the analyzer: a reflection common to the states, in the instrument, in the connector or in the mount, is absorbed by the two-port, and the quantities that the method obtains from the same reflection coefficients and that carry any error in them---the calibration factor and the efficiencies of the states at the floor, Fig.~\ref{fig:sp4}(b) and Fig.~\ref{fig:sp4sweep}(c)---are smooth in frequency, with the roughness of the other coaxial mounts. The period corresponds to an electrical length $c/2\Delta f=1.75$~cm in air (shorter in the dielectric of the capsule), longer than the thermistor capsule and comparable with the mount; a reflection at that distance would, however, be common to the states and absorbed by the two-port, as noted above. This is the best approximation found; no closed physical model of the mount reproduces it.

The specimen was then swept in the bridge resistance as the sensors of Section~\ref{sec:results}: 17 states, 100 to 220~$\Omega$ in steps of 10~$\Omega$ and 250, 300, 350 and 400~$\Omega$, on the same frequency grid as the first measurement and in one session of 40~min, and the correction was computed jointly for all of them (Fig.~\ref{fig:sp4sweep}). Three facts come out of the sweep. The modulation has one shape in frequency, the same in every state: a single function of frequency carries 96\,\% of the energy of the 17 curves of $s_i(f)$ about their band means (the first singular value of the matrix of those curves), with its zeros at 1.5, 5.7 and 9.9~GHz in all of them, and the phases $\theta_i(f)$ follow it. Its weight across the states (Fig.~\ref{fig:sp4sweep}(b)) changes sign at 140 and 300~$\Omega$, the two zeros of the per-state correction of the coaxial family (Section~\ref{sec:results}); the modulation is thus common to the states and is distributed among them by the selection~\eqref{eq:minnorm}, in the way the selection distributes everything that the two-port and the model leave: of its energy, 94\,\% lies at each frequency in the six-dimensional subspace that a change of a state-independent two-port spans on the impedances, $\delta\ln Z_i=(p+qZ_i+rZ_i^2)/Z_i$ with $p$, $q$, $r$ complex, and 6\,\% in its orthogonal complement. And in what depends on the state the mount is like the others: the part of the 17-state correction that no state-independent two-port can absorb is 0.05\,\% rms, against 0.03\,\% in coaxial specimen~2.

What the sweep leaves is therefore a correction whose state-dependent content is that of its family and whose modulation belongs to the specimen as a whole, the same in every state; its origin is not settled here. The quantities the method is for are obtained all the same: the effective efficiencies of the 17 states at the floor (Fig.~\ref{fig:sp4sweep}(c)) are smooth in $R_b$, within 1.2\,\% of $\eta_2$ over the whole range, with the operating state as the most efficient, as the floor requires, and their dispersion over the band is below 0.6\,\% in the hottest state and below 0.1\,\% from 160~$\Omega$ upwards. The specimen is kept in Table~\ref{tab:m} and left out of the per-state law~\eqref{eq:rplaw}, whose constants are not defined for its first-pass model.

\begin{figure}[!t]
\centering
\includegraphics[width=0.88\columnwidth]{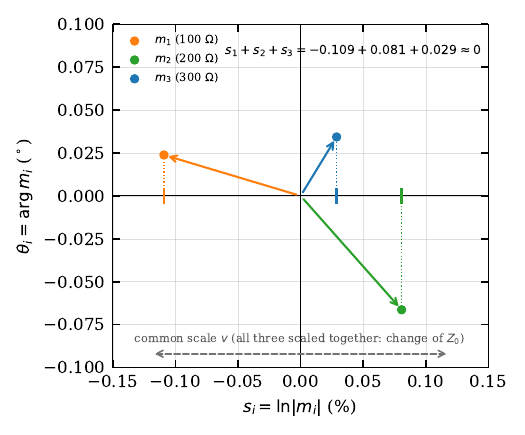}
\caption{The correction of coaxial specimen~2 at 5~GHz as three vectors in the plane of logarithmic modulus and phase, $Z'_i=m_iZ_{\mathrm{th},i}$. The common scale $v$, a displacement of the three impedances together along the horizontal axis, is a change of reference impedance that the two-port absorbs; the selection~\eqref{eq:minnorm} leaves no net component along it, $\sum_i s_i=0$. Axes in percent and degrees; the argument of Appendix~\ref{app:traceless} is in nepers and radians.}
\label{fig:appB}
\end{figure}
\section{Minimum-Norm Correction and the Traceless Property}
\label{app:traceless}
Let $x=(s_1,\dots,s_N,\theta_1,\dots,\theta_N)$ be the unknowns of~\eqref{eq:minnorm} for $N$ states, and $v=(1,\dots,1,0,\dots,0)$ the direction of a common logarithmic scale, $Z'_i\to e^{\epsilon}Z'_i$ for every $i$. Along $v$ the corrected impedances are scaled together, which is the same as referring the load plane to the impedance $e^{-\epsilon}Z_0$: a change of reference impedance. The two-port built from the triple absorbs it as a renormalization of its bead-side port, so that every state still satisfies $\Gamma_i=S(Z'_i)$; the efficiencies, which are ratios of powers, do not depend on the normalization, and neither does $F$ (both are invariant along $v$ to $10^{-12}$); and the gradient of $\sigma_{\max}$ with respect to $x$ has along $v$ a component of $4\times10^{-4}$ of its norm (coaxial specimen~2, 0.5--10~GHz). To first order, $v$ is therefore tangent to the constraint set of~\eqref{eq:minnorm} at the solution. The minimizer of $\|x\|^2$ on a constraint set is orthogonal to the tangent directions of that set at the minimizer, since otherwise a displacement along a tangent direction would lower the norm while keeping the constraints; hence $x\cdot v=\sum_i s_i$ is expected to vanish to that order, and so it is found: with $\|x\|$ of a few $10^{-3}$, $\sum_i s_i$ is below $10^{-5}$ at every frequency on the eleven sensors, for three states and for the swept sets of 23 and 31 states alike. A common phase, $Z'_i\to e^{j\phi}Z'_i$, is not a change of reference impedance---it is not a bilinear map of the load plane that keeps the measured $\Gamma_i$ with a passive two-port---and the sum of the $\theta_i$ is accordingly not constrained; it is small but not zero ($-0.05^\circ$ in the waveguide mounts). The property is thus one of the selection: the first pass fixes the constants of $Z_{\mathrm{th}}$ at the logarithmic center of the corrected impedances, and the correction distributes the departure from passivity among the states with no net scale. Fig.~\ref{fig:appB} shows the three correction vectors $(s_i,\theta_i)=(\ln|m_i|,\arg m_i)$ of coaxial specimen~2 at one frequency: the common scale $v$ is the horizontal direction, along which the three impedances would move together, and the three horizontal components cancel.

\section{The Passivity Floor}
\label{app:floor}
Let $S$ be the two-port of Fig.~\ref{fig:scheme}, with incident waves $a=(a_1,a_2)$ and reflected waves $b=Sa$, and let state $i$ terminate its bead port in the load $\Gamma_{L,i}$, so that $a_2=\Gamma_{L,i}b_2$ and $b_1=\Gamma_ia_1$. The power entering port~1 is $P_{\mathrm{in}}=|a_1|^2(1-|\Gamma_i|^2)$, the power absorbed by the load is $P_L=|b_2|^2(1-|\Gamma_{L,i}|^2)=\eta_iP_{\mathrm{in}}$, and the power dissipated in the two-port is their difference, $\|a\|^2-\|b\|^2=P_{\mathrm{in}}(1-\eta_i)$. Writing $\|a\|^2=|a_1|^2+|\Gamma_{L,i}|^2|b_2|^2=P_{\mathrm{in}}D_i$, with
\begin{equation}
D_i=\frac{1}{1-|\Gamma_i|^2}+\frac{\eta_i|\Gamma_{L,i}|^2}{1-|\Gamma_{L,i}|^2}\ \ge 1,
\label{eq:Di}
\end{equation}
the excitation of state $i$ gives $\|b\|^2/\|a\|^2=1-(1-\eta_i)/D_i$. The square of the spectral norm is the largest value of $\|b\|^2/\|a\|^2$ over all excitations, hence
\begin{equation}
\sigma_{\max}^2(S)\ \ge\ 1-\frac{1-\eta_i}{D_i}\ \ge\ \eta_i
\label{eq:floor}
\end{equation}
for every state $i$. Passivity, $\sigma_{\max}\le1$, states that no excitation extracts more power than it delivers; \eqref{eq:floor} is its counterpart from below: a two-port that delivers a fraction $\eta_i$ of the power it receives to a load cannot have a spectral norm smaller than $\sqrt{\eta_i}$, whatever its other elements. The efficiency of the operating state is fixed by the first pass through $K_f=\eta_2(1-|\Gamma_2|^2)$, so every two-port compatible with the data satisfies $\sigma_{\max}\ge\sqrt{\eta_2}$, and the two-port that attains the bound is the least active one compatible with the data: this is the floor of~\eqref{eq:minnorm}. Equality in~\eqref{eq:floor} requires the excitation of state~2 to be the principal singular vector of $S$ and $D_2=1$, i.e.\ $\Gamma_2=\Gamma_{L,2}=0$; otherwise $\sigma_{\max}$ exceeds $\sqrt{\eta_2}$ by at least $(1-\eta_2)(D_2-1)/2\sqrt{\eta_2}$ to first order in $D_2-1$, i.e.\ by approximately $(1-\eta_2)(|\Gamma_2|^2+\eta_2|\Gamma_{L,2}|^2)/2\sqrt{\eta_2}$, the term quoted in Section~\ref{sec:intro}, which with the values of the sensors is $2$--$4\times10^{-5}$ in the coaxial mounts and $3\times10^{-4}$ in band median in the waveguide mounts, up to $1.4\times10^{-3}$ at the lower band edge of specimen~4 where $|\Gamma_2|$ reaches 0.23; the offset of $10^{-3}$ added to $\sqrt{\eta_2}$ covers it everywhere but at those 23 of 715 points, and Section~\ref{sec:results} gives the effect of varying it. The bound does not depend on the normalization of port~1, which is that of the analyzer; $\Gamma_{L,i}$, and with it the attainability term, depend on the reference impedance of the bead port, which is that of the first pass of each family~\cite{greco2026coax,greco2026wg}.

\bibliographystyle{IEEEtran}
\bibliography{refs}

\begin{IEEEbiographynophoto}{Patricio Greco}
(Member, IEEE) received the degree in electronic engineering from the Universidad Tecnol\'ogica Nacional, Buenos Aires, Argentina, in 1999. He is an independent researcher in RF and microwave metrology, Buenos Aires, Argentina, and has run an aeronautical electronics workshop for 29 years. His research interests include the calibration of bolometric power sensors, thermistor modeling and vector network analyzer methods. He is a member of the IEEE Microwave Theory and Technology Society and of the IEEE Instrumentation and Measurement Society.
\end{IEEEbiographynophoto}

\end{document}